\documentclass[aps,pra,superscriptaddress,twocolumn, a4paper]{quantumarticle}

\pdfoutput=1

\usepackage{comment}
\usepackage[utf8]{inputenc}
\usepackage{amssymb}
\usepackage{amsmath}
\usepackage{amsfonts}
\usepackage{bm}
\usepackage{mathtools}
\usepackage{bbold}
\usepackage{upgreek}
\usepackage{xfrac}
\usepackage{soul}
\usepackage{bm}
\usepackage{comment}
\usepackage{xcolor}
\usepackage{mathtools}

\newtheorem{theorem}{Theorem}%[section]

\usepackage[colorlinks=true,citecolor=blue]{hyperref}
\hypersetup{colorlinks=true,citecolor=blue,linkcolor=blue,urlcolor=blue}
\usepackage{url}

\DeclareSymbolFont{matha}{OML}{txmi}{m}{it}% txfonts
\DeclareMathSymbol{\varv}{\mathord}{matha}{118}
\usepackage{bbm}
\begin{document}

\title{Protocols for Trainable and Differentiable Quantum Generative Modelling}

\author{Oleksandr Kyriienko}
\affiliation{Pasqal SAS, 2 av. Augustin Fresnel, 91120 Palaiseau, France}
\affiliation{Department of Physics and Astronomy, University of Exeter, Stocker Road, Exeter EX4 4QL, United Kingdom}

\author{Annie E. Paine}
\affiliation{Pasqal SAS, 2 av. Augustin Fresnel, 91120 Palaiseau, France}
\affiliation{Department of Physics and Astronomy, University of Exeter, Stocker Road, Exeter EX4 4QL, United Kingdom}

\author{Vincent E. Elfving}
\affiliation{Pasqal SAS, 2 av. Augustin Fresnel, 91120 Palaiseau, France}

\date{15 February 2022}

\begin{abstract}
We propose an approach for learning probability distributions as differentiable quantum circuits (DQC) that enable efficient quantum generative modelling (QGM) and synthetic data generation. Contrary to existing QGM approaches, we perform training of a DQC-based model, where data is encoded in a latent space with a phase feature map, followed by a variational quantum circuit. We then map the trained model to the bit basis using a fixed unitary transformation, coinciding with a quantum Fourier transform circuit in the simplest case. This allows fast sampling from parametrized distributions using a single-shot readout. Importantly, latent space training provides models that are automatically differentiable, and we show how samples from solutions of stochastic differential equations (SDEs) can be accessed by solving stationary and time-dependent Fokker-Planck equations with a quantum protocol. Finally, our approach opens a route to multidimensional generative modelling with qubit registers explicitly correlated via a (fixed) entangling layer. In this case quantum computers can offer advantage as efficient samplers, which perform complex inverse transform sampling enabled by the fundamental laws of quantum mechanics. On a technical side the advances are multiple, as we introduce the phase feature map, analyze its properties, and develop frequency-taming techniques that include qubit-wise training and feature map sparsification.
\end{abstract}

\maketitle

\section{Introduction}
Quantum computing (QC) promises to offer a computational advantage by meticulous usage of an exponentially large Hilbert space for qubit registers~\cite{Aaronson2013}. However, the use of QC is limited to specific tasks, as efficient solutions are only expected for some problem types~\cite{Aaronson2008}. One example corresponds to sampling from quantum states created by random entangling circuits~\cite{Terhal2004,Boixo2018}. This task lies at the heart of quantum supremacy experiments~\cite{Arute2019,JWPan2020,Wu2021,Zhu2021a}. While being computationally advantageous for producing samples (just need to send a `\emph{measure}' instruction), the considered distributions are not suitable for industrially relevant advantage~\cite{Harrow2017}, though may be helpful in studying related concept such as quantum chaos~\cite{Mi2021}. Finding a subset of problems with distributions which are both classically-intractable and industrially useful is an open challenge. Quantum generative modelling (QGM) aims to exploit trainable circuits that can prepare distributions as quantum states, for instance trying to match patterns from available data. Being a subject of the emerging field of quantum machine learning (QML)~\cite{Benedetti2019rev,Cerezo2020rev}, QGM utilizes the Born rule inherent to quantum mechanics~\cite{Born1926}. The goal is to represent a parametrized probability distribution $p_{\bm{\theta}}(x)$. It represents a probability to measure a bit string $x$ from a variational state $|\psi_{\bm{\theta}}\rangle$ parametrized by a vector of gate parameters $\bm{\theta}$~\cite{Han2018,Benedetti2019b}. For the simple case of pure states this reads $p_{\bm{\theta}}^{\mathrm{QCBM}}(x) = | \langle x | \psi_{\bm{\theta}}\rangle|^2$. This approach is the basis of quantum circuit Born machines (QCBMs)~\cite{Cheng2018} that learn models directly from samples of a target distribution $p_{\mathrm{target}}(x)$ using various loss functions~\cite{JGLiu2018,Coyle2020}. A similar approach is used for generating circuits in quantum generative adversarial networks (QGANs)~\cite{Zeng2019,Zoufal2019,Du2019,Lloyd2018,Dallaire2018}, where the training schedule corresponds to the minimax game. To date, QCBMs have been used for loading static distributions corresponding to bars-and-stripes dataset~\cite{JGLiu2018,Benedetti2019b}, learning datasets of correlated currency pairs~\cite{Coyle2021}, and digitized Gaussian and bimodal distributions~\cite{JGLiu2018}. QGANs were used for (reduced) MNIST datasets~\cite{Huang2020}, financial modelling~\cite{Zoufal2019}, learning pure states~\cite{Benedetti2019a}, and sampling particle traces~\cite{Chang2021}. 
While making a step towards sampling-based advantage, current QGM performance is largely limited even for idealized statevector simulators~\cite{JGLiu2018}. First, the described generators are difficult to train as they require matching all amplitudes for $N$-qubit registers and finding the corresponding state for some vector $\bm{\theta}$. Second, QCBM architecture is not automatically differentiable with respect to variable $x$, and QGAN differentiation leads to an ill-defined loss landscape \cite{Paine2021}. Thus, both have limited application for SDE solving. The latter would be hugely beneficial as differential constraints remove strong dependence on data, regularize models, and offer additional structure to learning (see quantum approach to adding differential constraints in \cite{Kyriienko2021,Knudsen2020} and physics-informed neural network architectures in classical machine learning \cite{yang2018physicsinformed,zhang2019learning,Raissi2018}). SDE-based sampling is also motivated by works in the financial sector where Monte-Carlo techniques are used. To date, various quantum protocols for associated PDEs has been considered, in many cases taking the perspective of real and imaginary time evolution~\cite{Gonzalez2021,Santosh2021,Kubo2021,Alghassi2021} or using amplitude amplification for tasks like option pricing~\cite{Rebentrost2018,Stamatopoulos2020,Martin2021,Goldman2021,Egger2020}. More broadly, the area of differential equations with quantum computers has been developing rapidly, starting from fault-tolerant QC oriented~\cite{Costa2019,Childs2020b,Linden2020,Childs2020a} to near-term and quantum-inspired protocols~\cite{Kyriienko2021,Knudsen2020,Lubasch2020,Wang2020,Garcia-Ripoll2021,Garcia-Molina2021,Goes2021}. Furthermore, differentiable distributions allow for the use of gradient ascent which enables extremal learning \cite{Patel2021}, with relevant applications in design/optimization tasks. 

We first note that the ability of differentiating generative models can be restored when using feature map encoding of continuous distributions \cite{Romero2021}, at the expense of multi-shot measurement to get a sample from QNNs. Second, the differential constraints at the sampling stage can be implemented using quantum quantile mechanics (QQM) \cite{Paine2021}, where a quantum circuit is trained to generate samples from SDEs and can be evolved in time, albeit with expectation-based sampling. Here, merging differentiability with fast sampling will offer both potential expressivity advantage and sampling advantage of QC.

In this work we develop a workflow for training of quantum generators that can be differentiated with respect to a continuous stochastic variable. For this, we separate the training and sampling stages of QGM. During the training stage we build a model in the latent space (taken as a phase space) enabled by the \emph{phase feature map}, followed by a variational circuit, and DQC-type readout. The sampling stage is then performed in the bit basis space enabled by the fixed unitary transformation (e.g. quantum Fourier transform), and followed by projective measurements for a sample-by-sample readout. The proposed workflow leads to \emph{differentiable} quantum generative models (DQGM \cite{dqgm_approach}), and is used for sampling from SDEs. Another consequence of training in the phase space is inherent model regularization, enforced by the proposed \emph{qubit-wise} learning, \emph{feature map sparsification}, and \emph{frequency-taming} techniques for circuit initialization based on Fourier series. Showing probability distribution (or generic function) loading into state amplitudes, we proceed to solve Fokker-Planck equations, giving access to time-series of the Ornstein-Uhlenbeck process. Finally, considering correlated registers where quantum correlations are included by entangling circuits \cite{Niu2021,Zhu2021b}, we discuss how classically hard multi-dimensional distributions can be automatically ``inverted'' by QCs, making a step towards a sampling advantage.

%--------------------------

\section{The approach}
Generative modelling concerns the process of drawing samples of a stochastic variable $X_t \sim p_{\bm{\theta},t}(x)$ from a trainable distribution with variational angles $\bm{\theta}$, which is also parametrized by $t$. Typically, we associate $t$ to time as a deterministic variable, which may enter explicitly (as an additional parameter) or implicitly encoded in $\bm{\theta}(t)$. We will use the notation $\bm{\theta},t$ throughout for both cases, and specify encoding where ambiguity may arise. In the generic quantum case the model can be constructed using Born's rule, $p_{\bm{\theta},t}(x) = \mathrm{tr} \{ |x\rangle \langle x| \hat{\rho}_{\bm{\theta},t} \}$, where samples $x$ corresponding to length-$N$ binary strings are readout from the density operator $\hat{\rho}_{\bm{\theta},t} = \mathcal{E}_{\bm{\theta},t}(\hat{\rho}_0)$ created by a parametrized completely positive trace-preserving (CPTP) map $\mathcal{E}_{\bm{\theta},t}$ from some initial density operator $\hat{\rho}_0$. 
%%%
\begin{figure}
\begin{center}
\includegraphics[width=1.0\linewidth]{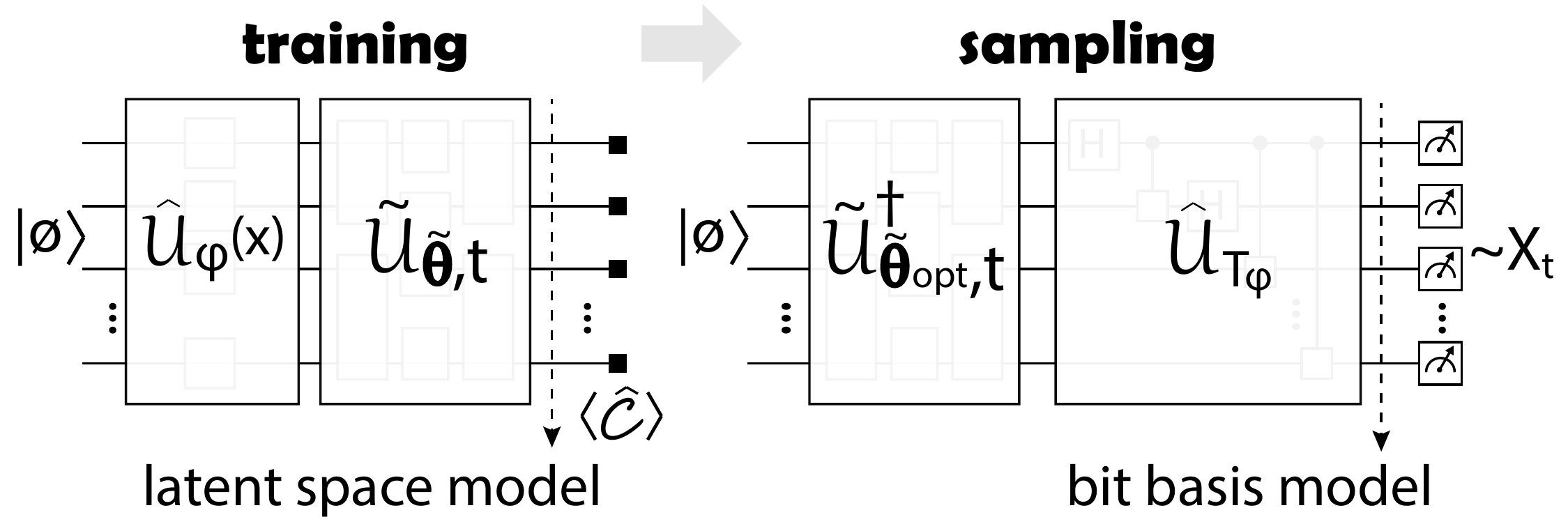}
\end{center}
\caption{\textbf{DQGM training and sampling.} At the training stage we use the latent space model representation, where the phase feature map directly follows by the variational circuit (and basis transformation circuits effectively cancel each other). At the sampling stage, we revert the trained variational circuit and map the model from the latent to the bit space, while the feature map and inverse basis transformation are treated as a part of the projective measurement, and are subsumed in a sampling process.}
\label{fig:workflow}
\end{figure}
%%%
The latter typically corresponds to the computational zero state $\hat{\rho}_0 = |{\o}\rangle \langle {\o}|$, where $|{\o}\rangle \equiv |0\rangle^{\otimes M}$ for $M \geq N$. In many cases unitary quantum channels are considered, $\mathcal{E}_{\bm{\theta},t}(\hat{\rho}_0) = \hat{\mathcal{U}}_{\bm{\theta},t} \hat{\rho}_0 \hat{\mathcal{U}}_{\bm{\theta},t}^\dagger$ with $M = N$ and $\hat{\mathcal{U}}_{\bm{\theta},t}$ is a generic parametrized unitary on $N$-qubit register. Note that when $\hat{\mathcal{U}}_{\bm{\theta}} \in \mathcal{SU}(2^N)$ in principle any state of the register can be prepared, and we call such a model maximally expressive. We recall that typically QCBM-style generative modelling relies on sample-based training of $p_{\bm{\theta},t}^{\mathrm{QCBM}}(x) = \mathrm{tr} \{ |x\rangle \langle x| \hat{\mathcal{U}}_{\bm{\theta},t} \hat{\rho}_0 \hat{\mathcal{U}}_{\bm{\theta},t}^\dagger \}$ at digital (i.e. integer, binary) values of $x$ only, and angles $\bm{\theta}$ are sought separately at different points of time $t$. The generic goal is minimizing a loss function $\mathcal{L}_{\bm{\theta},t}^{\mathrm{QCBM}} = \sum_{x = 0}^{2^N -1} \mathfrak{D}[p_{\mathrm{target}}(x,t), p_{\bm{\theta},t}^{\mathrm{QCBM}}(x)]$, for some distance measure $\mathfrak{D}[\cdot, \cdot]$. The optimization procedure gives the optimal angles $\bm{\theta}_{\mathrm{opt}} = \mathrm{argmin}_{\bm{\theta}} \big[ \mathcal{L}_{\bm{\theta},t}^{\mathrm{QCBM}} \big]$ at fixed $t$. In practice, this is achieved using data samples $x \in \mathcal{X}_{\mathrm{data}}$ (typically, from observations) and a proxy loss, corresponding to maximum mean discrepancy (MMD) \cite{JGLiu2018}, Stein discrepancy (SD) \cite{Coyle2020}, Kullback-Leibler divergence, as well as other types of f-divergences \cite{Leadbeater2021}. Once $p_{\bm{\theta},t}^{\mathrm{QCBM}}(x)$ is successfully trained, one can proceed directly to sampling from the same circuit.

We propose to act differently. We start by describing the protocol for generating computational states $\{ |x\rangle \}$ (each associated to binary strings $x \in \mathcal{B} = \{00..0, 10..0, \dots, 11..1 \}$). This can be achieved in two steps. First, a parametrized feature map creates a latent (phase) space representation of the variable $x$, $\hat{\rho}_{\tilde{x}} = \hat{\mathcal{U}}_{\varphi}(x) \hat{\rho}_0 \hat{\mathcal{U}}_{\varphi}^\dagger(x)$. For convenience, we call the corresponding circuit the phase feature map. For $\hat{\rho}_0 = |{\o}\rangle \langle {\o}|$ it reads
\begin{align}
\label{eq:Uphase}
\hat{\mathcal{U}}_{\varphi}(x) = \prod_{j=1}^N \left[ \hat{R}_j^z \left(\frac{2\pi x}{2^j}\right) \hat{H}_j \right],
\end{align}
where $\hat{R}_j^z(\phi) = \cos(\phi/2) \hat{\mathbb{1}}_j - i \sin(\phi/2) \hat{Z}_j$ is a single-qubit rotation and $\hat{H}_j$ is a Hadamard gate, acting at site $j$. Additionally, we include $\xi_j$ as (restricted) qubit-dependent coefficients that may be used for transforming (stretching or squeezing) the coordinate $x$. The circuit in Eq.~\eqref{eq:Uphase} maps an initial state into a superposition \emph{product} state $\hat{\rho}_{\tilde{x}} = |\widetilde{x}\rangle \langle \widetilde{x}|$ based on the latent state $|\widetilde{x}\rangle := \hat{\mathcal{U}}_{\varphi}(x) |{\o}\rangle$, which explicitly reads
\begin{align}
\label{eq:x_latent}
|\widetilde{x}\rangle = \frac{e^{-i\Phi/2}}{2^{N/2}} \bigotimes\limits_{j=1}^{N} \left( |0\rangle_j + \exp\left(-i \frac{2\pi x}{\xi_j 2^j} \right) |1\rangle_j \right),
\end{align}
where $\Phi = 2\pi (1-2^{-N})$ is an overall phase. Importantly, the phase space representation contains all computational basis states, which we can label by integers $\{x_\ell\} = \{0, 1, \dots 2^N - 1 \}$, and associated states are not entangled. Next, we apply the quantum circuit $\hat{\mathcal{U}}_{\mathrm{T}_\varphi}$ such that it transforms latent states $\{ |\widetilde{x}_{\ell}\rangle \}$ into binary states $\{ |x\rangle \}$ as a bijection. The subscript $\varphi$ highlights that the transformation circuit is designed for the specific feature map. The corresponding density operator $\hat{\rho}_x = \hat{\mathcal{U}}_{\mathrm{T}_{\varphi}} \hat{\rho}_{\tilde{x}} \hat{\mathcal{U}}_{\mathrm{T}_{\varphi}}^\dagger$ thus encodes the variable $x$ in the bit basis. We note that the simplest case for such transformation is for $\xi_j = 1~\forall~j$. In this case the mapping corresponds to the inverse quantum Fourier transform (QFT) circuit, $\hat{\mathcal{U}}_{\mathrm{T}_\varphi} = \hat{\mathcal{U}}_{\mathrm{QFT}}^\dagger$ \cite{NielsenChuang}, which consists of $O(N^2)$ gates (Hadamards and controlled-phase). Having generated the state $\hat{\rho}_x$ we proceed by applying a variational ansatz. We choose it in the form $\hat{\mathcal{W}}_{\tilde{\bm{\theta}},t} = \widetilde{\mathcal{U}}_{\tilde{\bm{\theta}},t} \hat{\mathcal{U}}_{\mathrm{T}_\varphi}^\dagger$, where with the tilde in $\widetilde{\bm{\theta}}$ and $\widetilde{\mathcal{U}}_{\tilde{\bm{\theta}},t}$ we highlight that the circuit structure and parametrization angles are different from QCBM. 
%such that at the training stage the QFT-like transforms cancel out and model keeps the latent space representation. 
Our strategy is building a differentiable quantum generative model (DQGM \cite{dqgm_approach}), fully in the latent space, $\widetilde{p}_{\tilde{\theta},t}(x) = \mathrm{tr}\{\hat{\mathcal{C}}_{\o} \widetilde{U}_{\tilde{\bm{\theta}},t} \hat{\rho}_{\tilde{x}} \widetilde{U}_{\tilde{\bm{\theta}},t}^\dagger\}$, with the cost (measurement) operator being $\hat{\mathcal{C}}_{\o} = \rho_0$. The model is trained to match the target distribution for $\widetilde{\theta}_{\mathrm{opt},t} = \mathrm{argmin}_{\tilde{\theta}} \sum_{x \in \mathcal{X}} \mathfrak{D}[p_{\mathrm{target}}(x,t), \widetilde{p}_{\tilde{\theta},t}(x)]$ for a grid $\mathcal{X}$ of real-valued $x \in [0,2^N-1)$ (or in other normalized interval), at given $t$. Note that due to training in the latent space the cost can be also a local operator~\cite{Cerezo2021}, or single-ancilla SWAP/Hadamard test for measuring the overlap. We then sample the trained model using projective measurements as $X_t \sim p_{\tilde{\bm{\theta}}_{\mathrm{opt}},t} = \mathrm{tr}\{ |x\rangle \langle x| \hat{\mathcal{U}}_{\mathrm{T}_\varphi} \widetilde{\mathcal{U}}_{\tilde{\bm{\theta}}_{\mathrm{opt}},t}^\dagger \hat{\rho}_0 \widetilde{\mathcal{U}}_{\tilde{\bm{\theta}}_{\mathrm{opt}},t} \hat{\mathcal{U}}_{\mathrm{T}_\varphi}^\dagger \}$ (see Fig.~\ref{fig:workflow}).
To show that we can sample the model successfully in the bit basis, let us formulate the connection between DQGM and QCBM in Theorem \ref{Theorem_1} below.
\begin{theorem}
\label{Theorem_1}
Probability distributions of binary samples $\{ X_t \}$ from maximally expressive QCBM at global optimum $\bm{\theta}_{\mathrm{opt}}$ and maximally expressive DQGM at global optimum $\widetilde{\bm{\theta}}_{\mathrm{opt}}$ are equivalent.
\end{theorem}
\noindent\textit{Proof.} Generative modelling from QCBM can be expressed as sampling from a generalized probability distribution
\begin{align}
\label{eq:gen_QCBM_1}
p_{\bm{\theta},t}^{\mathrm{gQCBM}}(x) &= \mathrm{tr} \{ |x\rangle \langle x| \hat{\mathcal{U}}_{\bm{\theta},t} \hat{\rho}_0 \hat{\mathcal{U}}_{\bm{\theta},t}^\dagger \} = \\
& = \mathrm{tr} \{ \hat{\mathcal{C}}_{\o} \hat{\mathcal{U}}_{\varphi}^\dagger(x) \hat{\mathcal{U}}_{\mathrm{T}_\varphi}^\dagger \hat{\mathcal{U}}_{\bm{\theta},t} \hat{\rho}_0 \hat{\mathcal{U}}_{\bm{\theta},t}^\dagger \hat{\mathcal{U}}_{\mathrm{T}_\varphi} \hat{\mathcal{U}}_{\varphi}(x)  \},
\label{eq:gen_QCBM_2}
\end{align}
where $\hat{\mathcal{U}}_{\varphi}^\dagger(x)$ corresponds to the phase feature map. At digital values of the variable Eq.~\eqref{eq:gen_QCBM_2} corresponds to $p_{\bm{\theta},t}^{\mathrm{QCBM}}(x)$, but extends QCBM to $x \in \mathbb{R}$. Note that in the intervals between digital points $\ell<x<\ell+1$ ($\ell=0,1,\dots,2^N-2$) the samples come from the superposition of neighboring states, $\propto \alpha |x_\ell\rangle + \beta |x_{\ell + 1}\rangle$ (with $x$-dependent complex coefficients $\alpha,\beta$), preserving sampling locality. The latent DQGM model can be rewritten as
\begin{align}
\label{eq:p_latent}
\widetilde{p}_{\tilde{\bm{\theta}},t}(x) = \mathrm{tr} \{ \hat{\rho}_{\tilde{x}} \widetilde{U}_{\tilde{\bm{\theta}},t}^\dagger \hat{\rho}_0 \widetilde{U}_{\tilde{\bm{\theta}},t} \} = \mathrm{tr} \{ |x\rangle \langle x| \hat{\mathcal{W}}_{\tilde{\bm{\theta}},t}^\dagger \hat{\rho}_0 \hat{\mathcal{W}}_{\tilde{\bm{\theta}},t} \},
\end{align}
directly following from cyclic properties of the trace and previously introduced definitions. Comparing models in Eq.~\eqref{eq:gen_QCBM_1} and Eq.~\eqref{eq:p_latent}, and given that quantum states $\hat{\mathcal{U}}_{\bm{\theta},t} \hat{\rho}_0 \hat{\mathcal{U}}_{\bm{\theta},t}^\dagger$ and $\hat{\mathcal{W}}_{\tilde{\bm{\theta}},t}^\dagger \hat{\rho}_0 \hat{\mathcal{W}}_{\tilde{\bm{\theta}},t}$ are trained to match the same target distribution, for maximally expressive circuits $\hat{\mathcal{U}}_{\bm{\theta},t}, \widetilde{U}_{\tilde{\bm{\theta}},t} \in \mathcal{SU}(2^N)$ the probability distributions match at the global optimum, $p_{\bm{\theta}_{\mathrm{opt}},t}^{\mathrm{gQCBM}}(x) = \widetilde{p}_{\tilde{\bm{\theta}}_{\mathrm{opt}},t}(x)$. This follows from the fact that both circuits are \emph{in principle} capable of expressing any state (quasi-distribution)~\cite{PreskillNotes}, where $\hat{\mathcal{W}}_{\tilde{\bm{\theta}},t}$ can absorb a fixed transformation by re-adjusting the angles, and both aim to prepare the same optimal state.\vspace{1mm}

While we show that the two approaches are equivalent during the sampling stage, the two models are vastly different during the training stage. For the QCBM and its generalization in Eq.~\eqref{eq:gen_QCBM_2} the sampling and training settings are the same. They require a variational state to match bit string probabilities already in training. This basis may work better for peaked or discrete distributions (like bars-and-stripes), but challenging for smooth functions. %This is a challenging setting and calls for deep variational circuits~(see discussion in \cite{Romero2021}). 
%We formalize the claim in the remark below.
%\begin{remark}
%Generalized QCBM circuits are not trainable.
%\end{remark}
%The reason behind that training of $p_{\bm{\theta}}^{\mathrm{gQCBM}}(x)$ is inhibited by dataset-induced barren plateaus that cannot be remedied by (potentially) trainable variational circuits \cite{Thanasilp2021}.
%
For the DQGM we only require training of the latent model, where a superposition product state is obtained from $x$-parametrized single qubit rotations (spans all $O(2^N)$ amplitudes) and needs a certain overlap with a variational state (with a support of the same size). Intuitively, this task is easier to achieve, and we substantiate the claim later when conducting numerical experiments. As DQGM and QCBM originate from the same phase feature map, they have the same \emph{model capacity} --- spectrum characterized by exponentially large number of frequencies (considered in the next subsection). At the same time, DQGM has better \emph{model expressivity} in terms of access to Fourier coefficients for relevant low-frequency components, thanks to the (non-variational) unitary transformation $\hat{\mathcal{U}}_{\mathrm{T}_\varphi}$ that can remove a part of the training complexity.\vspace{1mm}

\subsubsection*{Model differentiation and constrained training from stochastic differential equations}

One of the important consequences of the proposed approach is the possibility for differentiating a constructed quantum model. This can be done by using quantum automatic differentiation (AD) applied to the phase feature map~\cite{Kyriienko2021}. Note that as we use the latent model in training, we can apply differential constraints already at this stage. Only once trained we proceed to sampling. Let us discuss examples where such physics (or finance/biology/chemistry) constraints are important. Consider a stochastic differential equation written as~\cite{OksendalBook}
\begin{align}
\label{eq:SDE}
    d X_t = f(X_t, t) dt + g(X_t, t) dW_t,
\end{align}
where $dW_t$ is a standard Wiener process, $X_t$ is time-dependent stochastic variable, and $f(\cdot)$, $g(\cdot)$ are some scalar functions typically referred as drift and diffusion. For any SDE in the form \eqref{eq:SDE} we can write an equation of motion for the probability distribution. This can correspond to a Fokker-Planck equation (FPE) or a Kolmogorov backward equation (KBE) \cite{Bogachev2015}, written for the time-dependent probability distribution function $p(x,t)$ of the stochastic variable $X_t$. More generally, the evolution can be described by the Feynman-Kac formula~\cite{Alghassi2021}. Importantly, once we learn the $p(x,t)$ in the domain of interest $t \in \mathcal{T}$, in-principle we can obtain stochastic trajectories (samples from time-incremented distributions), offering full generative modelling of time-series. Normally, given only access to a function $p(x,t)$, generating samples requires a costly inversion procedure (or equivalent), and is challenging for multidimensional problems. For the quantum generative models it requires learning $t$-parametrized DQGM at different times, giving direct access to fast sampling. Below we sketch the workflow, and provide more details when considering examples in the Results section.

The stochastic problem \eqref{eq:SDE} can be approached from a data-driven perspective, where we first learn a representation of the steady state from available samples. This is highly relevant also from the point of view of model discovery \cite{Heim2021}, as drift and diffusion coefficients may not be immediately known. Setting the loss function for DQGM as  $\mathcal{L}_{\bm{\theta},t_0}^{\mathrm{data}} = \sum_{x \in \mathcal{X}} \mathfrak{D}[p_{\mathrm{target}}(x, t_0), \widetilde{p}_{\bm{\theta}, t_0}(x)]$, we can learn a distribution at a point of time $t_0$. 

Now, let us comment on two possible ways of encoding the time variable. First, time $t$ can be embedded explicitly. One option is to use a $t$-dependent feature map for parameterizing the model. For instance, we employed it successfully in DQC-based quantum function propagation~\cite{Paine2021}. In this case, it is convenient to use an identity-valued feature map at $t_0$, and learn to adjust angles as $t$ deviates from $t_0$. Second, explicit encoding of time can take a polynomial of $t$ (or even a feed-forward neural network), with $\bm{\theta}$'s being trainable coefficients. In this case, $t = t_0$ training can be performed for zeroth degree term, and adjusting remaining coefficients at other times. Finally, we can also assume an implicit dependence of variational coefficients $\bm{\theta}(t)$ on time. In this case, we learn to represent data at $t_0$ with parameters $\bm{\theta}(t_0)$, and then demand that each point of time the distribution satisfies differential constraints for a PDE in question. This leads to model-dependent updates of variational parameters $\bm{\theta}(t+\Delta t) \xleftarrow{\gamma} \bm{\theta}(t)$ (with an update rule $\gamma$), thus evolving the model in discrete time \cite{YDu2021}. Below, we show how to introduce model-dependent differential constraints, and training or evolving DQGM in both explicit and implicit manner. We note both are physics-informed, and represent a step forward from static sample generation. %In the Results section we also detail a procedure for evolving the distribution in the implicit form.

Given the SDE in \eqref{eq:SDE}, the evolution of associated $p(x,t)$ requires solving a PDE either forward or backward in time. The former case corresponds to solving the Fokker-Planck equation (corresponding to the Kolmogorov forward equation). A generic FPE can be written as 
\begin{align}
\label{eq:FPE_general}
\frac{\partial }{\partial t} p(x,t) = &-\frac{\partial }{\partial x} \left[ f(x,t) p(x,t) \right] \\ \notag & + \frac{1}{2} \frac{\partial^2 }{\partial x^2} \left[ g^2(x,t) p(x,t) \right],
\end{align}
and we evolve the system towards the stationary state at $t_{\mathrm{s}} > t$ from some initial distribution. The stationary distribution of FPE then satisfies the second-order differential equation
\begin{align}
\label{eq:FPE_ss}
\mathrm{FPE}(p,x, t_{\mathrm{s}};f,g):= &-\frac{d }{d x} \left[ f(x,t_{\mathrm{s}}) p(x,t_{\mathrm{s}}) \right] \\ \notag &+ \frac{1}{2} \frac{d^2 }{d x^2} \left[ g^2(x,t_{\mathrm{s}}) p(x,t_{\mathrm{s}}) \right] = 0,
\end{align}
and we call the corresponding differential constraint on the distribution the FPE differential operator. Specifically, we can substitute $p(x,t_\mathrm{s})$ with $\widetilde{p}_{\bm{\theta},t_\mathrm{s}}(x)$ and train a quantum generative model to respect the FPE constraint assigning the differential loss $\mathcal{L}_{\bm{\theta},t_\mathrm{s}}^{\mathrm{diff}} = \sum_{x \in \mathcal{X}} \mathfrak{D}[0, \mathrm{FPE}(\widetilde{p}_{\bm{\theta},t_\mathrm{s}},x; f,g)]$, such that it remains true for all $x$. We note that this inherently regularizes the model, and in particular leads to improved derivative matching, highly relevant for studying tails of distributions and dynamics.

Next, we note that we can train a quantum model to represent the PDF at some point of time $t_0$, using data as a snapshot during evolution. Then, the full PDE and associated differential constraints are used to propagate it in the $t_0 < t < t_\mathrm{s}$ interval reaching the steady state at $t_\mathrm{s}$. Specifically, we can write the differential loss based on the difference of the RHS and the LHS of the FPE, which we call the dynamical FPE differential operator $\mathrm{DFPE}(p,x,t; f,g)$. The loss dictates that our model minimizes $\mathcal{L}_{\bm{\theta}}^{\mathrm{evol}} = \sum_{x,t \in \mathcal{T}\times \mathcal{X}} \mathfrak{D}[0, \mathrm{DFPE}(\widetilde{p}_{\bm{\theta},t},x; f, g)]$, and we assume explicit time embedding. Then the workflow for evolving differentiable quantum generative models has a style similar to PINN/DQC workflow~\cite{Kyriienko2021}. Once done, the model can be sampled within the trained region, and generalized in between the points.

Alternatively, we can use an evolutionary approach for updating circuit parameters~\cite{YDu2021}. In this case, the time-derivative of our model $\partial \widetilde{p}_{\bm{\theta},t}(x) / \partial t$ can be re-expressed using a chain rule as $(\partial \widetilde{p}_{\bm{\theta},t}(x) / \partial \bm{\theta}) (\partial \bm{\theta}/\partial t)$. The differential constraints in space and time then require that a vector of updates satisfies $\gamma = (\mathbf{J}^{T}\cdot\mathbf{J})^{-1}\cdot\mathbf{J}^{T}\cdot\mathbf{F}$, where $\mathbf{F}$ is a vector corresponding to differential operator $\mathrm{FPE}(\widetilde{p}_{\bm{\theta},t},x; f, g)$ evaluated at the grid $x \in \mathcal{X}$. The matrix $\mathbf{J}$ is the Jacobian for our model evaluated at $x \in \mathcal{X}$, each having $|\bm{\theta}|$ entries. The update can be performed using a simple Euler's forward update $\bm{\theta}(t+\Delta t) = \bm{\theta}(t) + \Delta t \gamma$, where $\Delta t$ is a time step, and we stress that $\gamma$ is recalculated as we ``march'' over the grid of times. Going beyond linear updates, more sophisticated schemes (e.g. Runge-Kutta) can be employed.

Finally, we can evolve the probability distribution using the Kolmogorov backward equation (KBE), where the goal is to study the dynamics at times prior to the steady state. Let us define $\tau < t_{\mathrm{s}}$ as a backward time. A generic KBE associated to the SDE \eqref{eq:SDE} reads
\begin{align}
\label{eq:KBE_general}
- \frac{\partial }{\partial \tau} p(x,\tau) = f(x,\tau) \frac{\partial }{\partial x} p(x,\tau) + \frac{g^2(x,\tau)}{2} \frac{\partial^2 }{\partial x^2} p(x,\tau).
\end{align}
It is convenient to set a starting point $\tau = t_{\mathrm{s}}$ and find $p(x, \tau < t_{\mathrm{s}})$ backward in time, discovering (and sampling) the model at earlier times. All steps discussed before apply here as well. 

%{\color{blue}NB: Need to introduce $t$-dependence into DQGM, any suggestions where?} {\color{red}VE: In-principle, in QQM the time dimension and space dimensions can be treated on equal footing, i.e. baking in a multi-dimensional aspect is similar to making a t-dependence. However, here if you think about it more closely to the application perspective, we don't really want to sample from p(x,t) and get a sample like (X,T) out. We want to query p(x,t) at t=T and that is different; i.e. we don't want a sample T}

Once we define the setting for solving problems based on SDE/PDE, we need to specify how to differentiate the proposed model (something that is not possible with QCBM/QGAN architectures). In the next subsection, where we analyse the phase feature map, we will also show how to read out $x$ derivatives of DQGM. While this can be done through the parameter shift rule~\cite{Mitarai2018,Schuld2019} and generalizations~\cite{Kyriienko2021b}, can be readout exactly and more efficiently by avoiding the regular parameter shift rule.\vspace{1mm}

\subsection{Phase feature map analysis} We note that by construction the latent space probability distribution $\widetilde{p}_{\tilde{\bm{\theta}}}(x)$ corresponds to a parametrized quantum circuit with feature map encoding \cite{Schuld2019a,Goto2020,Schuld2021,Caro2021}, and can be analyzed by studying associated Fourier series (for brevity, we omit $t$ dependence in this subsection). We proceed to analyse the model capacity of the phase feature map $\hat{\mathcal{U}}_{\varphi}(x)$. While Chebyshev series are available with additional variable transformations~\cite{Kyriienko2021}, for the phase map with homogeneous $\{ \xi_j = 1\}_{j=1}^{N}$ we remain in Fourier space. Specifically, we define capacity as the number of modes (frequencies) that are \emph{in principle} available in the model. This is determined by the spectral properties of the generator of the feature map, $\hat{G}:~\hat{\mathcal{U}}_{\varphi}(x) = \exp(-i x \hat{G} /2)$. We note that parametrized quantum circuits can generally represent a function (model) as 
\begin{align}
\label{eq:model_Fourier}
f_{\bm{\theta}}(x) = \sum\limits_{\omega \in \Omega} c_{\omega,\bm{\theta}} e^{i \omega x},
\end{align}
where the spectrum of frequencies $\Omega$ represent all possible \emph{differences} of eigenvalues of $\hat{G}$, and $c_{\omega,\bm{\theta}}$ are $\bm{\theta}$-dependent coefficients associated to each frequency~\cite{Schuld2021,Kyriienko2021b}. The important properties of the spectrum are that it includes zero frequency, pairs of equal-magnitude positive and negative frequencies, and coefficients obey $c_{\omega} = c_{-\omega}^*$ leading to real-valued models (as expected from an expectation value). While the analysis can proceed by studying the generator of the phase map, here we derive model capacity explicitly from the latent state written in Eq.~\eqref{eq:x_latent}. Let us define the phase for each qubit rotation as $\varphi_j := 2\pi/(2^j \xi_j)$. The $N$-qubit superposition state $|\widetilde{x}\rangle$ has an equal overlap with all computational basis states, $|\langle x | \widetilde{x}\rangle|^2 = 1/2^N~\forall~x\in\mathcal{B}$, but each individual contribution comes with a different phase (sum of individual $\varphi_j$'s). Expanding the tensor product in Eq.~\eqref{eq:x_latent} we see that the computational zero state $|{\o}\rangle$ has a phase of zero, by convention. Next, there are $N$ states with single excitations, $|j\rangle := e^{i\varphi_j x} \hat{X}_j |{\o}\rangle$, each with their phase exponentially decreasing from the highest ($\varphi_1 = 2\pi/2$) to lowest ($\varphi_N = 2\pi/2^N$) as qubit number increases. Next, we have $N(N-1)/2$ states with double excitations, $|jj'\rangle := e^{i(\varphi_j + \varphi_{j'}) x} \hat{X}_{j} \hat{X}_{j'} |{\o}\rangle$, with corresponding phases of a sum of contributions. In general, there are $N!/m!(N-m)!$ states with $m$ excitations (and sums of $m$ phases), culminating with a fully excited state $|\mathbb{1}\rangle := e^{i\Phi} \hat{X}^{\otimes N} |{\o}\rangle$, with $\Phi = \sum_j \varphi_j = 2\pi (2^N - 1)/2^N$. We collect sum of phases associated to bit basis states $\{ |x_\ell\rangle \}$, calling them frequencies $\{ \nu_{\ell} \} = \{ 2 \pi \ell / 2^N  \}_{\ell = 0}^{2^N - 1}$ at this point. We note that the latent state can be rewritten in a simple form $|\widetilde{x}\rangle = (e^{-i\Phi/2} / 2^{N/2}) \sum_{\ell=0}^{2^N - 1} e^{i \nu_\ell x} |x_\ell\rangle$. Next, we proceed to construct the model itself as in Eq.~\eqref{eq:p_latent}, which comes from the overlap (squared) of the latent feature state with an ansatz-prepared state, $\hat{\mathcal{U}}_{\bm{\theta}} |{\o}\rangle = \sum_{\ell = 0}^{2^N-1} a_{\ell,\bm{\theta}} |x_{\ell}\rangle$ (hereafter we simplify the notation by removing tildes where appropriate). The latent space probability distribution then reads
\begin{align}
\label{eq:p_model_exp}
&\widetilde{p}_{\bm{\theta}}(x) = \frac{1}{2^N} \sum\limits_{\ell,\ell'=0}^{2^N - 1} a_{\ell,\bm{\theta}}^* a_{\ell',\bm{\theta}} e^{i (\nu_\ell - \nu_{\ell'}) x} = \\ \notag
& =  \frac{1}{2^N} + \frac{1}{2^{N-1}} \sum\limits_{\ell > \ell'} \bigg\{ \mathrm{Re}\{ a_{\ell,\bm{\theta}}^* a_{\ell',\bm{\theta}}  \} \cos[(\nu_\ell - \nu_{\ell'}) x] \\ \notag &- \mathrm{Im}\{ a_{\ell,\bm{\theta}}^* a_{\ell',\bm{\theta}} \} \sin[(\nu_\ell - \nu_{\ell'}) x] \bigg\},
\end{align}
where in the second and third line of Eq.~\eqref{eq:p_model_exp} we split the double sum to show real and imaginary part of the $\bm{\theta}$-dependent density operator elements $a_{\ell,\bm{\theta}}^* a_{\ell',\bm{\theta}}$, and account for quantum state normalization. We recall that frequencies $\{ \nu_\ell \}$ are simply integer multiples of the smallest (`base') frequency $2\pi/2^N$ defined by the register size. Looking at the differences of $\{\nu_\ell - \nu_{\ell'}\}_{\ell,\ell'=0}^{2^N - 1}$ we observe that the model in Eq.~\eqref{eq:p_model_exp} corresponds to Eq.~\eqref{eq:model_Fourier} with $\omega \in \Omega = \{ 0, \pm 1, \pm 2, ..., \pm (2^N -1)\} \times 2\pi/2^N $, where multiplicity for each frequency decreases as $2^N - \ell$, $\ell = 0, 1, \cdots, 2^N -1$, and we just need to collect associated coefficients $c_{\omega,\bm{\theta}}$ for each $\omega$. We thus see that the spectral properties of the phase feature map and associated latent model establish its capacity of exponential size with $(2^N -1)$ non-zero frequencies, and the same degree (times the base frequency) \cite{Schuld2021}.

Given the analysis above, we draw several conclusions that are highly important for the successful training of quantum generative models. We list them below.\vspace{1mm}
%A family of models accessible by DQGM is that of trigonometric polynomials with exponentially many frequencies and constrained variationally-controlled coefficients. It is an arduous challenge to find optimal variational angles in this setting from randomly initialized ansatz acting on the full register.    

\noindent 1. Both DGQM and QCBM have $O(2^N)$ model capacity, but have different model expressivity in terms of coefficients $\{ c_{\omega,\bm{\theta}} \}$. As variational unitary circuits have limited depth due to trainability, the performance will widely vary depending on typically accessible model coefficients for the given ansatz \cite{Schuld2021}. The exponential capacity can then be seen as a problem for certain distributions (see discussion in Ref.~\cite{Caro2021}), as highly-oscillatoric terms will lead to overfitting and corrupt derivatives when solving differential equations.\vspace{1mm}
%%%
\begin{figure}
\begin{center}
\includegraphics[width=1.0\linewidth]{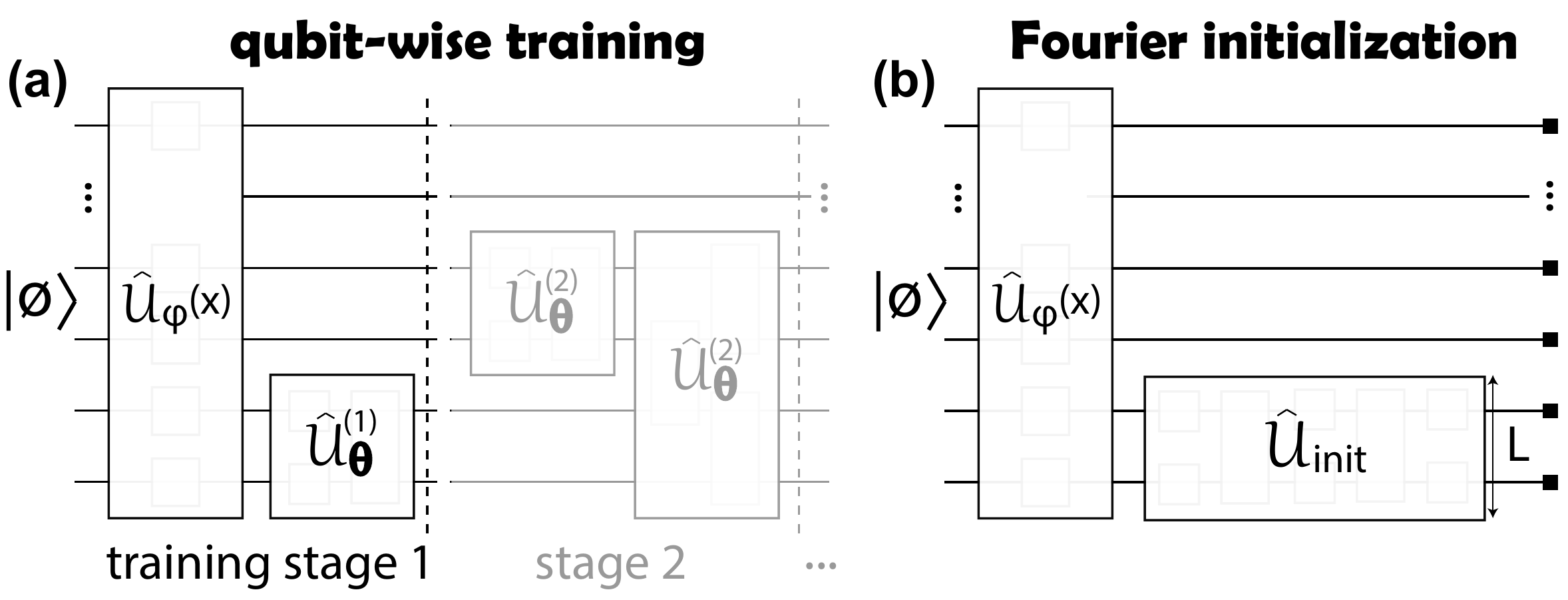}
\end{center}
\caption{\textbf{Frequency-taming techniques.} (a) Qubit-wise training, where variational circuit is first trained to adjust low frequency part of a model (stage 1). In the second stage we keep $\hat{\mathcal{U}}_{\bm{\theta}}^{(1)}$ fixed, and train the higher frequency components with $\hat{\mathcal{U}}_{\bm{\theta}}^{(2)}$, also correlating it with the lower frequency register. This continues until sufficient accuracy. The final optimization run is for the full circuit and register. (b) For the Fourier initialization we first find classical Fourier series for a distribution of interest with $(2^L -1) \sim \mathrm{poly}(N)$ frequencies, and use $\hat{\mathcal{U}}_{\mathrm{init}}$ to prepare the corresponding state.}
\label{fig:f-taming}
\end{figure}
%%%

\noindent 2. In latent space there is a clear separation between high and low frequency parts of the model, corresponding to qubits with small and large $j$. This suggests that DGQM can be trained to adjust mostly low frequency components while keeping high frequency components intact, and use the full register for sampling. This is the core of qubit-wise training described in the next subsection. We note that such an approach does not hold for QCBMs.\vspace{1mm}

\noindent 3. A family of models accessible by DQGM is that of trigonometric polynomials with exponentially many frequencies and constrained variationally-controlled coefficients. In cases where a smooth probability distribution is modelled it may suffice to train only the low-frequency part of the register $L < N$ chosen such that $2^L \sim \mathrm{poly}(N)$. This allows for classical Fourier (cosine/sine) series to be used for probability distribution modelling and/or differential equation solving. The quantum model then requires $O(\mathrm{poly}(N))$ depth circuit as an instruction for creating the state $\hat{\rho}_{\bm{\theta}}$ that matches this series. In this case we can initialize the system close to a predicted solution (performing Fourier series initialization), but still getting sampling advantage for the full register and only using the variational state preparation for inducing further correlations.\vspace{1mm}

\noindent 4. The structure of the phase map is quite peculiar --- unlike product and tower feature maps~\cite{Kyriienko2021}, where phases of $x$-dependent rotations are either qubit-independent or have a prefactor of $j$, the phase feature map has $\sim 2^{-j}$ scaling. Thus, for the same capacity of the phase and product feature maps, the latter has higher expressivity as more qubits and wider variational circuits are used. We address this issue by proposing several feature map `frequency-taming' techniques in the next section. 

%------------

\subsection{Frequency-taming techniques}

In this subsection we describe several strategies that can be used for DGQM training. Specifically, we exploit the knowledge of latent space to perform training in several stages and provide means of regularizing trained generative models.

\subsubsection{Qubit-wise learning} As one of the frequency taming techniques for DQGM training we consider splitting the ansatz into lower and higher frequency parts. We call this qubit-wise learning, similarly to the layer-wise learning in classical and quantum machine learning~\cite{Skolik2021}. We sketch the procedure in Fig.~\ref{fig:f-taming}(a), where training is broken into stages. First, the goal is to get the base frequencies right for the model, and qubits $j = N, N-1, \dots$ are trained. Next, we save quasi-optimal angles for the first cycle of optimization, and proceed to include higher frequencies (qubits with smaller $j$). It is also important to correlate the registers, possibly with a tailored ansatz, and this question is a matter of future research. Finally, when all quasi-optimal angles are found, we perform training for the full register. %The approach will be demonstrated later with numerical experiments.

\subsubsection{Fourier initialization} One of the common problems affecting machine learning models is initialization that leads to local minima, and prohibits finding high-quality models. In Ref.~\cite{Paine2021} we have shown that initialization with low-degree polynomial (truncated Chebyshev series) can vastly reduce number of optimization epochs. Here, we propose to use the structure of the quantum model in Eq.~\eqref{eq:p_model_exp}, and match coefficients for all frequencies $\omega \in \Omega$ by preparing a suitable quantum state $\hat{\mathcal{U}}_{\mathrm{init}} |0\rangle^{\otimes L} = \sum_{\ell = 0}^{2^L -1} a_{\ell,\mathrm{init}} |x_\ell\rangle$ [Fig.~\ref{fig:f-taming}(b)]. Note that the preparation circuit can be exponentially deep in $L$ (see circuit construction in Ref.~\cite{Mottonen2005}), but since we only care about $\mathrm{poly}(N)$ frequencies we choose $L \ll N$, suggesting that this is a feasible step for cases where limited expressivity suffices, but fast sampling is needed for dataset augmentation (and specifically relevant for multi-dimensional distributions).
%%%
\begin{figure}
\begin{center}
\includegraphics[width=1.0\linewidth]{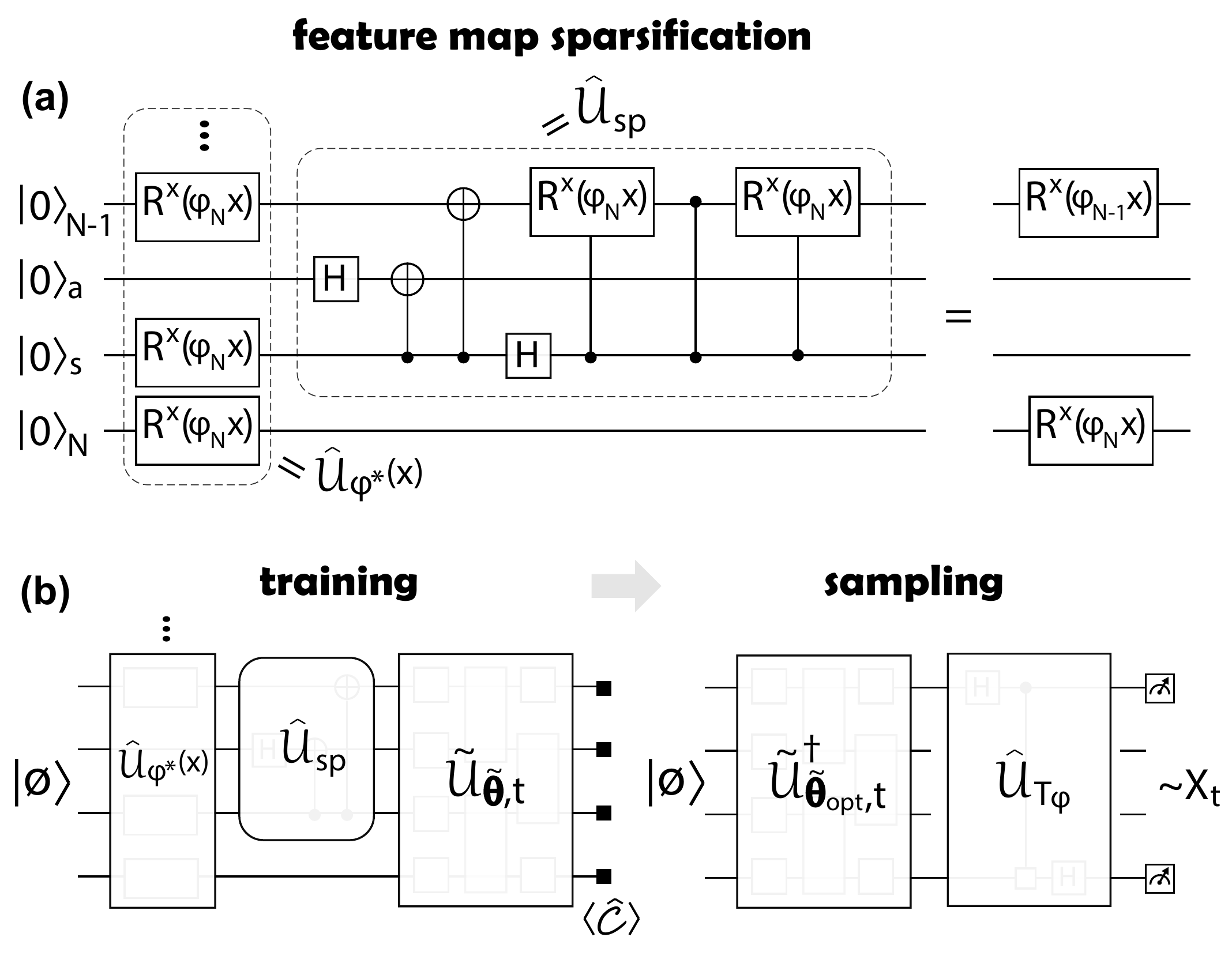}
\end{center}
\caption{\textbf{Feature map sparsification.} (a) Low-frequency part of the phase feature map, where the rotation gate from the seed qubit ($s$) is teleported to the register qubit $N-1$, which stores the second lowest frequency. Higher-frequency sparsifications can be constructed in the similar way, with varying split in frequencies (degree of sparsification). (b) Training and sampling stages for the sparsified phase map, where the variational circuit acts on all qubits including seeds and ancillas, while during sampling only the $N$-qubit register is transformed and measured. Again, only lowest frequencies are shown.}
\label{fig:sparsification}
\end{figure}
%%%

\subsubsection{Feature map sparsification} As we noted before, one of the desirable features when working with a feature map of exponential capacity is the possibility to control coefficients for different frequencies. For example, the comparison of serial and product feature maps in Ref.~\cite{Schuld2021} has shown that for the same model capacity the product feature map had better expressivity as already with a layer of rotations one has independent control over multiple coefficients, unlike the serial case. For the phase feature map we are in the situation where feature map rotations are concatenations of base frequency rotations, and no variational control of the model is allowed at that stage --- to enable sampling we cannot simply change the feature map as it is an integral part of the measurement circuit. We overcome this issue by proposing the strategy for spreading the features over larger number of qubits, which we name the feature map sparsification strategy.

The idea relies on the fact that we can concatenate two circuits if we use a modified quantum gate teleportation circuit \cite{Gottesman1999}. Note that we have chosen to work in the X Pauli basis for simplicity as the spectrum of the models is the same, and given that $\hat{H} \hat{Z} = \hat{X} \hat{H}$ we simply append an extra layer of Hadamards to the transformation circuit $\hat{\mathcal{U}}_{\mathrm{T}_\varphi}$. We show the sparsification workflow in Fig.~\ref{fig:sparsification}. Concentrating on lowest frequencies, we observe that the second-to-last qubit in the feature map shall be in the $\hat{R}^x(\varphi_{N-1} x)|0\rangle_{N-1}$ state, and $\varphi_{N-1} = 2 \varphi_{N}$.  We can prepare the same state by merging two rotations from different qubits. We take a \emph{seed} state as $\hat{R}^x(\varphi_{N} x)|0\rangle_{\mathrm{s}}$ [labelled as $s$ in Fig.~\ref{fig:sparsification}(a)]. Using a Bell state with an ancilla qubit, we can teleport the state from the seed to the register qubit, such that an additional $R^x(\varphi_{N}x)$ gate is applied. Note that the process can be made deterministic if we add an $x$-dependent correction circuit. In this case sparsification is performed by the unitary gate $\hat{\mathcal{U}}_{\mathrm{sp}}$, and circuit identity in Fig.~\ref{fig:sparsification}(a) holds.

It is important to stress that we can use sparsification during the training stage, where all qubits (including ancillas and seeds) are trained to match the model --- this does not change the frequencies, but increases expressivity. %We derive an analytical model for the simplest case to support this claim (see Appendix). 
Next, during the sampling stage we then use the trained model, but only sample qubits from the state register on which the transformation circuit acts.

\subsubsection{Phase map differentiation}

We recall that the DQGM model is built as
\begin{align}
\label{eq:DQGM_model}
\widetilde{p}_{\theta}(x) = \mathrm{tr}\{\hat{\mathcal{C}}_{\o} \hat{\mathcal{U}}_{\bm{\theta}} \hat{\mathcal{U}}_{\varphi}(x) \hat{\rho}_{0} \hat{\mathcal{U}}_{\varphi}^\dagger(x) \hat{\mathcal{U}}_{\bm{\theta}}^\dagger\}.
\end{align}
Our goal is to evaluate $d \widetilde{p}_{\theta}(x) / dx$ analytically (i.e. in a bias-free manner). For this, first observe that 
\begin{align}
\label{eq:dUdx}
\frac{d \hat{\mathcal{U}}_{\varphi}(x)}{dx} = -i \hat{M}_x \hat{\mathcal{U}}_{\varphi}(x),
\end{align}
where we introduce the operator $\hat{M}_x := \pi \sum_{j=1}^{N} \hat{X}_j/ 2^j$ as the generator of the phase map (again, we use the X Pauli basis for convenience). We note that it commutes with the map trivially, $[\hat{M}_x, \hat{\mathcal{U}}_{\varphi}(x)] = 0~\forall~x$. %Although $\hat{M}_x$ is non-unitary, we can still imagine that it is applied to the computational zero as $\hat{M}_x |{\o}\rangle \propto \sum_{j=1}^{N} |j\rangle /2^j$, and generates a `one-hot' state with single excitation. We can assign the normalization prefactor as $\mathcal{N}_M$ s.t. $(\mathcal{N}_M^{-1}) \hat{M}_x |{\o}\rangle$ is a valid quantum state. 
We recall that $\hat{\mathcal{C}}_{\o} = \hat{\rho}_{0}$.

Now we proceed to differentiating the full model, which gives
\begin{align}
\label{eq:dpdx}
\frac{d \widetilde{p}_{\theta}(x)}{dx} &= i\mathrm{tr}\{\hat{\rho}_{0} \hat{\mathcal{U}}_{\bm{\theta}} \hat{\mathcal{U}}_{\varphi}(x) \hat{M}_x \hat{\rho}_{0} \hat{\mathcal{U}}_{\varphi}^\dagger(x) \hat{\mathcal{U}}_{\bm{\theta}}^\dagger\}  \\ \notag
&- i\mathrm{tr}\{\hat{\rho}_{0} \hat{\mathcal{U}}_{\bm{\theta}} \hat{\mathcal{U}}_{\varphi}(x) \hat{\rho}_{0} \hat{M}_x \hat{\mathcal{U}}_{\varphi}^\dagger(x) \hat{\mathcal{U}}_{\bm{\theta}}^\dagger\},
\end{align}
where we change the order in which $\hat{M}_x$ acts on $\hat{\rho}_0$. We observe that the corresponding measurement of two overlaps can be combined into the measurement of the expectation value
\begin{align}
\label{eq:dpdx_exp}
\frac{d \widetilde{p}_{\theta}(x)}{dx} = \mathrm{tr}\{(\delta_1 \hat{\mathcal{C}}) \hat{\mathcal{U}}_{\bm{\theta}} \hat{\mathcal{U}}_{\varphi}(x) \hat{\rho}_{0} \hat{\mathcal{U}}_{\varphi}^\dagger(x) \hat{\mathcal{U}}_{\bm{\theta}}^\dagger\},
\end{align}
where we defined a differential cost operator $\delta_1 \hat{\mathcal{C}} := i \hat{M}_x \hat{\mathcal{C}}_{{\o}} - i \hat{\mathcal{C}}_{{\o}} \hat{M}_x$. Note that the result is valid for both global and local cost operators. For instance, for the global cost the modified differential cost operator can be rewritten as 
\begin{align}
\label{eq:diff_cost}
\delta_1 \hat{\mathcal{C}} = \pi \sum\limits_{j=1}^N \frac{1}{2^j} \hat{Y}_{j} \otimes |{\o}\rangle_{\bar{j}} \langle {\o}|, 
\end{align}
and the state $|{\o}\rangle_{\bar{j}}$ simply means that we are in zero for the register of $N-1$ qubits, apart from the $j$-th one.
We see that we need $N$ evaluations of this expectation. This is an improvement over the $2N$ evaluations for the parameter shift rule. By analysing the commutators in $\hat{\delta \mathcal{C}}$, that correspond to SWAP-like operators, we may possibly do better, and this is a question for future research. 

Similarly, we can write a second-order derivative for the quantum probability distribution. For this, we can differentiate the expression in \eqref{eq:diff_cost}, and observe that $d^2 \widetilde{p}_{\theta}(x)/dx^2$ can be written as an expectation value
\begin{align}
\label{eq:d2pdx2_exp}
\frac{d^2 \widetilde{p}_{\theta}(x)}{dx^2} = \mathrm{tr}\{ (\delta_2 \hat{\mathcal{C}}) \hat{\mathcal{U}}_{\bm{\theta}} \hat{\mathcal{U}}_{\varphi}(x) \hat{\rho}_{0} \hat{\mathcal{U}}_{\varphi}^\dagger(x) \hat{\mathcal{U}}_{\bm{\theta}}^\dagger\},
\end{align}
where we introduce another Hermitian operator
\begin{align}
\label{eq:diff2_cost}
\delta_2 \hat{\mathcal{C}} := 2 \hat{M}_x \hat{\mathcal{C}}_{{\o}} \hat{M}_x - \hat{M}_x \hat{\mathcal{C}}_{{\o}} - \hat{\mathcal{C}}_{{\o}} \hat{M}_x, 
\end{align}
which can be decomposed into $O(N^2)$ non-commuting terms and measured separately.

\subsection{Preparing multidimensional correlated distributions}

It is unlikely that sampling from a single univariate distribution using a quantum computer gives a computational advantage over using a classical computer. In the end, for most practical cases we can use --- for example --- a finite-degree polynomial approximation. This is commonly used in financial analysis. 
%%%
\begin{figure}[t]
\begin{center}
\includegraphics[width=1.0\linewidth]{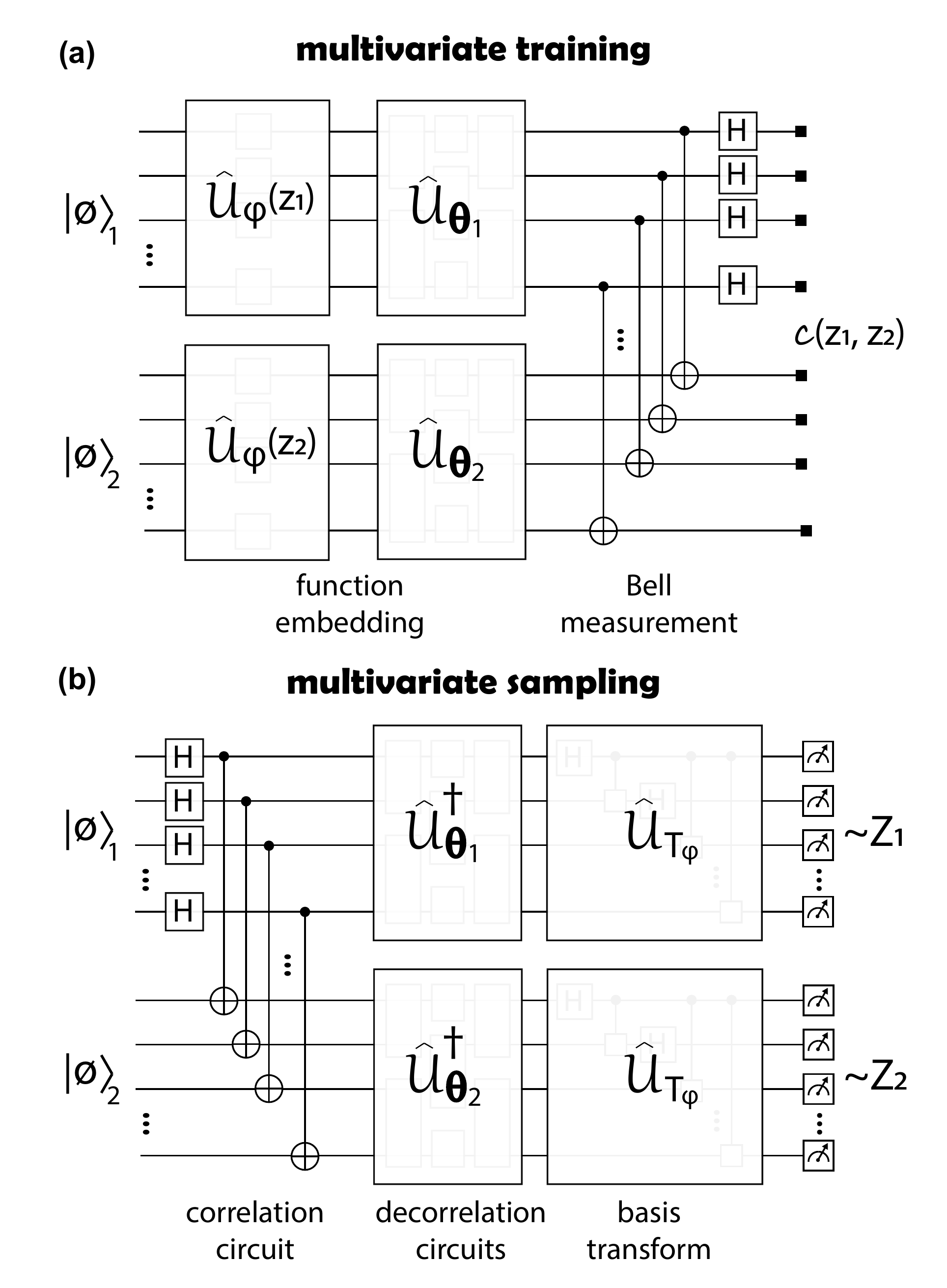}
\end{center}
\caption{\textbf{Multivariate quantum  generative models based on copulas.} (a) A model is trained to represent a copula dependence for latent variables, where the correlation between registers is included as series of Bell measurements. (b) The trained model is sampled in the bit basis starting from the cluster state that is transformed by variational circuits.}
\label{fig:correlations}
\end{figure}
%%%
However, when working with multivariate (multidimensional) distributions, sampling becomes complicated. This prompts us to consider problems comprising of a $D$-dimensional vector of stochastic variables $\bm{X} = (X_1, X_2, \dots, X_D)$. The underlying probability distribution corresponds to $p(\bm{x})$ with $\bm{x} = (x_1, x_2, \cdots, x_D)$, and often it is convenient to work with a multivariate cumulative distribution function $F(\bm{x})$. If the distributions are not correlated we can do inverse sampling assuming that the multivariate CDF factorizes into a product of marginal distributions, $F_{\mathrm{simple}}(\bm{x}) = F_1(x_1) \cdot F_2(x_2)\dots F_D(x_D)$, and the same is true for the probability density function. This means, even though we consider multivariate distributions, the simulation can be parallelized efficiently following the univariate case. However, for correlated variables this decoupling procedure is not valid. Classical simulation of multivariate distributions and corresponding generative modelling is generally difficult. Potential approaches include delayed rejection adaptive Metropolis algorithm, and the state-of-the-art protocols based on a tensor train decomposition \cite{Dolgov2020}. In general, they assume truncation of correlations, and the full generative modelling requires including fine structure, at large computational cost.

A way for including correlations between stochastic variables can be provided by quantum hardware, as quantum systems are good at correlating subsystems. Recently, generative modelling was shown to benefit from correlation, and specifically entanglement \cite{Niu2021}. One way to think about it is simply consider a joint register for the vector of variables $\bm{x}$. However, in this case we are left with a QCBM-type problem of enlarged size, and training for large $D$ can become prohibitive. A more subtle way corresponds to including correlations by encoding copulas into quantum hardware, as recently proposed in \cite{Zhu2021b}.

The concept of copula was developed to yield multivariate sampling by correlating latent variables, while keeping the sampling procedure individual to each variable. Imagine a bivariate distribution such that two stochastic variables $X_1$ and $X_2$ are distributed normally, but are in fact correlated. The correlation for normal distributions can be accounted using a covariance matrix, which grows with the dimension $D$. Thus, accounting for correlations again becomes challenging for generic $D$-dimensional distributions. However, this problem can be resolved by introducing a copula -- a function that links marginal distributions of different variables \cite{Nelsen2006}. Copulas absorb correlations between variables while being agnostic to the type of marginal distribution. Specifically, following Sklar's theorem we write a copula $C[\bm{v}]$ acting on some vector $\bm{v}$ as a function
\begin{align}
\label{eq:CDF_copula}
F(\bm{x}) = C[F_1(x_1), F_2(x_2), \dots, F_D(x_D)],
\end{align}
which links marginals into a full multivariate CDF. Similarly, a copula density function $c[\bm{z}]$ for the latent variable vector $\bm{z}$ is defined as
\begin{align}
\label{eq:copula_density}
c[\bm{x}] = c[F_1(x_1), \dots, F_D(x_D)] p_1(x_1) \cdot \dots p_D(x_D).
\end{align}
A useful property of copulas is that by generating a vector of samples from the copula as $\bm{Z} = (Z_1, Z_2, \dots, Z_D) \sim c$, we can transform them into samples of the original multivariate distribution as~\cite{Nelsen2006}
\begin{align}
\label{eq:copula_sampling}
\bm{X} = (Q_1(Z_1), Q_2(Z_2), \dots, Q_D(Z_D)),
\end{align}
where $Q_j(Z_j)$ are marginal quantile functions (inverted CDFs) for distribution of $j$-th stochastic variable. Here, we stress that copula produces correlations at the level of latent variables, as used in the inverse sampling~\cite{Paine2021}. It represents a modified PDF that deviates from a uniform multivariate distribution, and thus correlates the outcomes for multivariate PDF sampling.

Since the copulas capture correlations only, while having flat marginals, they can be modelled by entangled states \cite{Zhu2021b}. Namely, the correlations can be introduced using a quantum circuit of finite depth that is applied \emph{prior} to separate variational registers (see Fig.~\ref{fig:correlations}). Yet, when we link $D$ registers, even for tractable $N$-wide individual distributions, we are left with $D \cdot N$ qubits that are maximally entangled, in the logical sense. As we form a cluster state, this requires the bond dimension to go up, preventing efficient classical simulation. This is the setting in which we expect to get an advantage in quantum generative modelling.
%%%
\begin{figure*}[t!]
\begin{center}
\includegraphics[width=1.0\linewidth]{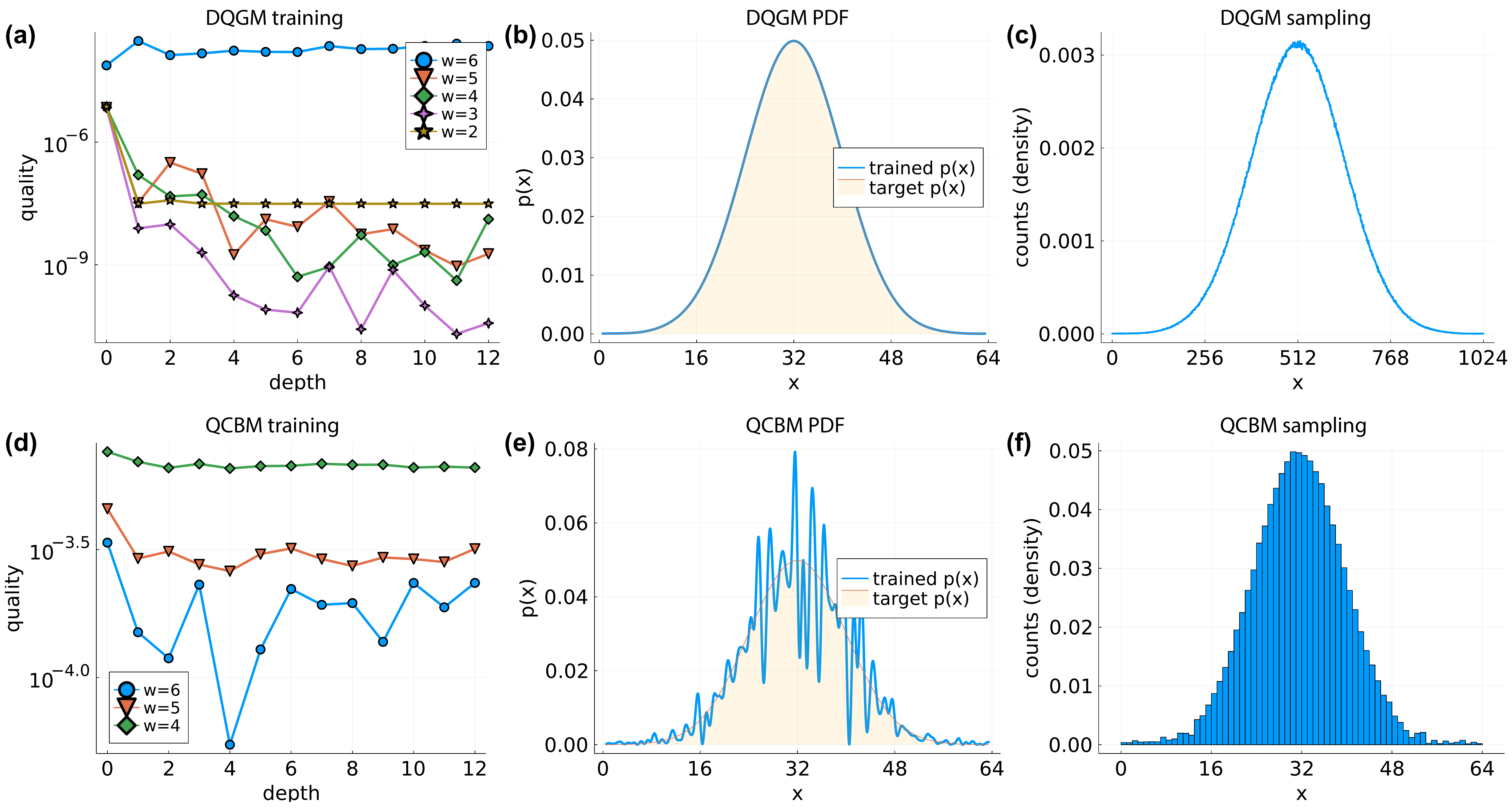}
\end{center}
\caption{\textbf{DQGM and QCBM comparison.} (a) MSE loss for DQGM trained at different depths and widths, showing quality of solution on the generalized grid. This corresponds to the quality metric, where smaller numbers (deviation) means higher quality. (b) PDF from the DQGM training at $d=4$ and $w = 3$. (c) Sampled probability distribution from transformed DQGM using $N=10$ qubits and $10^7$ samples at the readout. (d) Quality metric based on the MSE loss for the generalized QCBM trained at different depth and width. (e) Best model for QCBM shown for $d=4$ and $w=N$. (d) QCBM samples from $N=6$ qubits and $10^6$ shots.}
\label{fig:model-comparison}
\end{figure*}
%%%

We propose to build a quantum generative model for copulas, expressing it as a function of latent variables encoded using the phase feature map. The corresponding circuits for quantum copula modelling are shown in Fig.~\ref{fig:correlations}(a, b). First, the copula PDF is constructed as a function of variables $\bm{z}$ using the feature map encoding. We note that both DQGM and generalized QCBM models can be built. In the former case one needs to think in terms of frequencies, and in the latter case one shall think in terms of bit strings. The model is then constructed by first applying variational circuits on separate registers, then followed by the Bell circuit measurement and expectation of the cost operator $\hat{\mathcal{C}}_{{\o}}$ (global or local) [see Fig.~\ref{fig:correlations}(a)]. Intriguingly, this setting is similar to learning from data that has shown a great promise recently \cite{Huang2021c}, and uncovering the relation between two subjects is an interesting avenue for the future research. Once we trained the model for copula, we can revert the circuit, and read out samples in the transformed basis for DQGM [Fig.~\ref{fig:correlations}(b)]. Note that the probability density function remains the same. For the generalized QCBM, we note that $\hat{U}_{\mathrm{T}_{\varphi}}$ is a part of training, while being absent in the sampling stage.

We highlight that while building a quantum generative model for copulas, one can build powerful intuition about processes in the system. First, we observe that by generating a cluster state and using identity operators instead of variational circuits one enforces maximally correlated samples of $c(z_1, z_2)$. This in turn leads to strong correlation for samples $\bm{X} \sim p(\bm{x})$. However, by performing local operations on registers of separate variables one can effectively decorrelate their samples in the copula space, and thus in the space of multivariate PDF samples. We elaborate on this point in the Results section considering an example based on a Gaussian copula for bivariate distributions \cite{Meyer2013}.

Furthermore, the importance of representing a copula as a differentiable quantum model comes from the fact that for many stochastic processes (for instance, in financial modelling) certain copulas are shown to perform well, and represent an excellent starting point~\cite{Nelsen2006}. Going beyond learning from data, one can use  knowledge of differential constraints when learning copulas. This creates inherent regularization and helps capturing properties specific to the process. For instance, the system of Fokker-Planck equations formulated for a copula PDF, and used as a differential constraint, may offer an edge when training copula circuits \cite{Choe2013}.% We hope that providing the intuition for copulas in terms of QGM, and introducing a hybrid learning protocols from data and differential constraints, may help scalable sampling in higher dimensions.

\section{Results}
To test the proposed protocols, we conduct several numerical experiments. For this, we choose the Ornstein-Uhlenbeck process as an underling model~\cite{OksendalBook}. Being a starting point for the Hull-White and Vasicek models, Ornstein-Uhlenbeck SDE helps with, amongst others, modelling currency exchange rates \cite{Coyle2021}. First, we test the approach on learning a static distribution. Second, we introduce differential constraints and solve the steady-state FPE for OU. Third, we evolve the learnt solution in time, specifically solving the time-dependent FPE for OU using the implicit time embedding. Finally, we present results for multivariate sampling with quantum copula models.

%----------------------------

\subsection{Learning generative models}
We start with representing a probability density function (PDF) by DQGM circuits, with consequent sampling, and additionally compare it to the generalized QCBM architecture to highlight the differences in training. We choose the target distribution that corresponds to a normal process (Ornstein-Uhlenbeck being one example). The corresponding PDF reads
\begin{align}
\label{eq:p_target}
p_{\mathrm{target}}(x) = \frac{1}{\sqrt{2\pi \sigma_0^2}} \exp\left[-\frac{(x-\mu_0)^2}{2\sigma_0^2}\right],
\end{align}
where $\mu_0$ is a mean and $\sigma_0^2$ is a variance. We note that to be able to load a PDF in a quantum register suitable parameter scale should be chosen. Namely, $\mu_0$ and $\sigma_0$ are chosen such that the  probability can be potentially stored in a register of $N$ qubits with $x \in [0,2^N-1)$ and $0 \leq p(x) \leq 1$. We choose the mean square error (MSE) as a loss, which is normalized by the number of samples at which distributions are compared. As a testing ansatz for simplicity we use a hardware efficient ansatz (HEA) \cite{Kandala2017} with alternating $\mathcal{SU}(2)$ rotations and CNOT-based entangling layers. Specifically, we compose a variational circuit of $d$ layers and width $w$. Here, $d=0$ corresponds to single $\mathcal{SU}(2)$ layer (for instance, decomposed into X-Z-X parametrized rotations), followed by $d$ repetitions of CNOTs on odd/even sublattices and $\mathcal{SU}(2)$ layers. The parameter $w$ defines on how many qubits the variational ansatz acts, starting from the bottom one (lowest frequency). For instance, $w=3$ for $N=6$ register means we only use qubits $j = 4, 5, 6$, and act with an identity on the rest. Variation is performed using gradient-based Adam optimizer, and we use Julia's Yao package as a simulator \cite{Luo2019}.
%%%
\begin{figure}
\begin{center}
\includegraphics[width=0.85\linewidth]{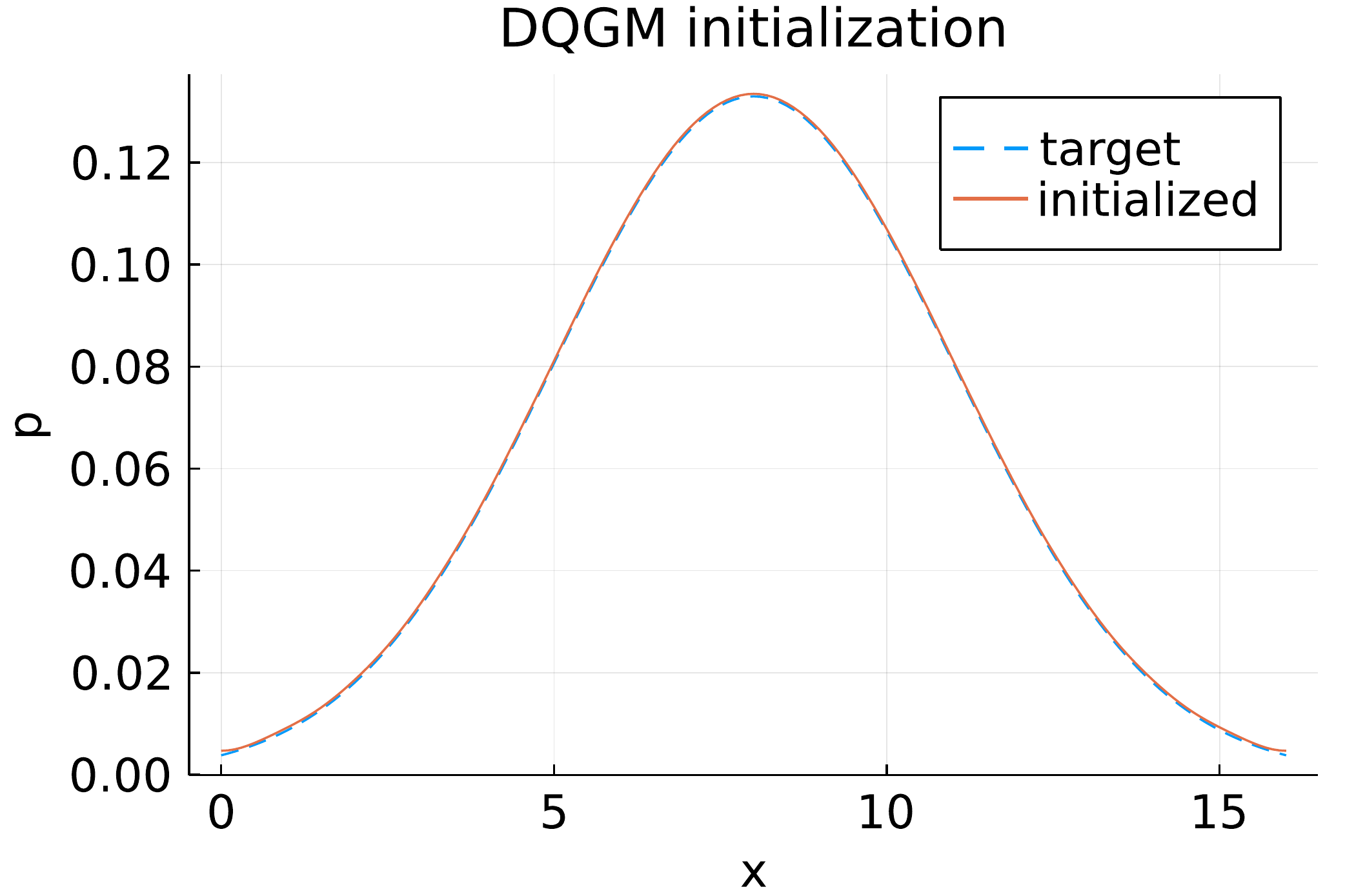}
\end{center}
\caption{\textbf{Fourier initialization of DQGM.} We use a cosine expansion and initialize the circuit for $L = N$, reaching high quality solutions and exploiting the full spectrum of $N=4$ DQGM.}
\label{fig:initialization}
\end{figure}
%%%

We start by considering a target distribution with $N = 6$ qubits. We set the mean to $\mu_0 = 32$ and the standard deviation of $\sigma_0 = 8$. The training grid is set up to include all integer points of $x$, and we use a thousand of epochs. The training is performed for varying depth and width. We test the performance of both DQGM and generalized QCBM for modelling the target as well as providing samples. As a metric, we plot the quality of the solution, represented by the MSE loss evaluated for twenty times more points (referred as a generalized grid). The results are shown in Fig.~\ref{fig:model-comparison}. 
In Fig.~\ref{fig:model-comparison}(a) we show the quality of solution for DQGM at the end of training. We observe that at full width training the model contains exponential number of frequencies, limiting the performance due to large `out-of-sample' error. At the same time, with smaller width we can capture the target distribution using lower frequency components, and reach high quality solutions. While the performance is likely to be model dependent, we observe that the optimal solution requires choosing a suitable combination of $w$ and $d$. As an example of trained PDF we pick $d=4$ and highest-performing width of $w=3$. This can be seen as a simplest instance of qubit-wise learning, and generally highlight the relevance of frequency-taming. The trained DQGM closely follows the target model at all points [see Fig.~\ref{fig:model-comparison}(b)]. We then apply the basis transformation and sample our model with the extended register of $M = 10$ qubits. The histogram is shown in Fig.~\ref{fig:model-comparison}(c), where $10^7$ shots are used, and we normalize bins over the total number of samples.

Next, we consider the performance of generalized QCBM for the same problem. The results for $d$ and $w$ scanning are depicted in Fig.~\ref{fig:model-comparison}(d). As encoding assumes transformations on bitstrings, smaller $w$ circuits do not perform well, and $w=N$ is required, as expected. We note that the presence of high frequencies in the model and absence of regularization that limits high frequency components generally impacts the QCBM's performance. The instance with the best quality is shown in Fig.~\ref{fig:model-comparison}(e). While overall the shape represents the distribution well, high-frequency components impact the model quality as it does not generalize. The impact on solving differential equations based on such a model will be tremendous. This can traced directly to the exponential capacity of the phase feature map, and the absence of simple frequency-taming. One option for regularization here is including more points during training, but this comes at the price of training on dense grids. Finally, we show the sampling from generalized QCBM in Fig.~\ref{fig:model-comparison}(f). The histogram qualitatively matches with the target, as requested by optimization loss.
%%%
\begin{figure*}
\begin{center}
\includegraphics[width=1.0\linewidth]{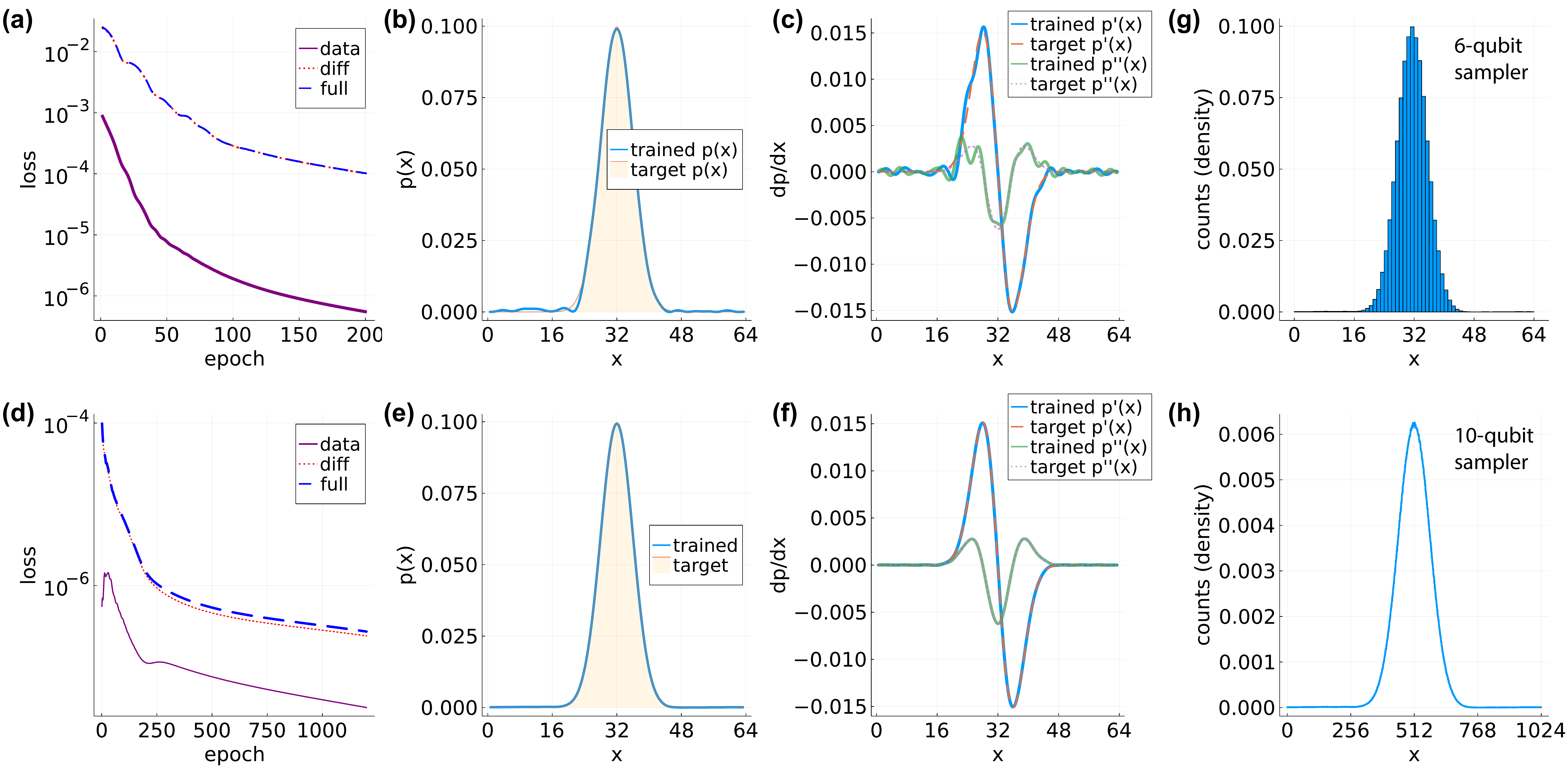}
\end{center}
\caption{\textbf{DQGM trained to sample from the Ornstein-Uhlenbeck process by matching the steady state of FPE.} (a) History of the data training, showing the data loss which is used for optimization. Differential loss (labelled as `diff') and the full weighted loss are plotted for comparison. (b) Probability distribution $p(x)$ from the data-trained DQGM, where small number of epochs is used. (c) Derivatives of the model trained on data. (d) History of DQGM training with differential constraints (stationary FPE), where the full weighted loss is used for optimization, and the other two loss functions are plotted for comparison. (e) The probability distribution function from DQGM trained on the full loss. (f) Derivatives of the generative model based on FPE constraints. (g) Normalized sampling histogram for $N=6$ DQGM trained using FPE differential constraints, where $10^6$ shots are measured. (h) Normalized sampling histogram from an extended 10-qubit register.}
\label{fig:FPE}
\end{figure*}
%%%

Following the use of the variational approach, we have also implemented the initialization procedure. In this case the target distribution is expanded in cosine series for $4$ qubits, such that the coefficients of the preparation state are known. Using a $\mathcal{SO}(2^4)$ circuit that can create an arbitrary real-amplitude state, we efficiently utilize all frequencies. The resulting PDF is shown in Fig.~\ref{fig:initialization}. We note that initialization may be required in cases where we want to off-load part of job from the variational procedure.

%For the initialization we reach the MSE loss of $2.354\times10^{-7}$, using all available frequencies and a state prepared by the fixed circuit. The corresponding plot is show in Fig.~\ref{fig:initialization}. The solution shows that high quality results can be obtained already at this system size. Moreover, since DQGM is formulated in terms of frequencies, we can easily increase the width of the system, getting more samples of the same quality.

\subsection{Solving stationary Fokker-Planck equations}

We proceed to introduce  differential constraints, where together with learning from data by minimizing $\mathcal{L}_{\bm{\theta}}^{\mathrm{data}}$, we wish to minimize $\mathcal{L}_{\bm{\theta}}^{\mathrm{diff}}$ coming from the FPE differential operator. While the data-based learning does not require knowing the model parameters per se, the SDE/PDE/ODE learning does depend on the model parameters introduced by the drift and diffusion terms. We again choose the Ornstein-Uhlenbeck process as it lies at the core of many financial models. SDE of the OU process corresponds to static drift and diffusion terms, and reads
\begin{align}
\label{eq:OU-SDE}
d X_t = -\nu (X_t - \mu) dt + \sigma dW_t ,
\end{align}
where $\mu$, $\sigma$, and $\nu$ are model parameters, which can be discovered while learning from data. Using Eq.~\eqref{eq:FPE_ss} we can see that at the steady state FPE for OU corresponds to
\begin{align}
\label{eq:FPE_ss_OU}
\nu p(x, t_{\mathrm{s}}) + \nu (x - \mu) \frac{d}{dx}p(x, t_{\mathrm{s}}) + \frac{\sigma^2}{2} \frac{d^2}{d x^2} p(x, t_{\mathrm{s}}) = 0.
\end{align}
Notably, when starting from some initial mean, we arrive to $\mu$ as a new mean in the steady state (at the rate of $\nu$), and the variance $\sigma^2/2\nu$. It is convenient to set $\nu = 1$, assuming that time is in units of $\nu^{-1}$.

In the following we assume that OU reached the steady state, and learn the corresponding distribution from the differential constraints. The workflow is as follows. First, we choose SDE/FPE parameters as $\mu_0 = 32$ and the variance of $\sigma_0^2 = 32$. The quantum model is set up with $N=6$ qubits, $d=4$ and $w=3$ as suggested by previously performed depth scanning. We set up three different loss functions to track the performance during training. The first two correspond to the data loss $\mathcal{L}_{\bm{\theta},t_\mathrm{s}}^{\mathrm{data}}$ and the differential loss for static FPE, $\mathcal{L}_{\bm{\theta},t_\mathrm{s}}^{\mathrm{diff}}$, as described in the second section. The third loss, which we call the full loss, is then taken as a weighted average of data and differential contribution,  $\mathcal{L}_{\bm{\theta},t_\mathrm{s}}^{\mathrm{full}} = \mathcal{L}_{\bm{\theta},t_\mathrm{s}}^{\mathrm{data}} + \eta \mathcal{L}_{\bm{\theta},t_\mathrm{s}}^{\mathrm{diff}}$, where coefficient $\eta$ controls the weight of FPE constrained (this is generally needed as the two may be imbalanced even when normalized over the grid). We perform DQGM training in two stages. At first, our goal is learning the initial condition of FPE, where the gradient descent is performed on $\mathcal{L}_{\bm{\theta},t_\mathrm{s}}^{\mathrm{data}}$. We deliberately choose a coarser grid with $32$ points of $x$ and $200$ epochs, simulating imperfect training conditions (i.e. when knowledge of probability distribution is not available, and data is noisy). The results are shown in Fig.~\ref{fig:FPE}(a-c). Looking at the history, the training loss goes down promptly, yet we observe a large separation between the data and diff loss contributions [Fig.~\ref{fig:FPE}(a)]. In Fig.~\ref{fig:FPE}(b) we show the corresponding PDF which captures the data well. Yet when plotting derivatives of the target model and DQGM in Fig.~\ref{fig:FPE}(c) significant deviations are visible. The latter can impact predictions when considering out-of-sample examples.
In the second stage we turn on the differential loss, and the full loss with equal contributions ($\eta= 1$). We use angles from the data training. Smaller learning rates are used to avoid jumping far from the previously found valley in a landscape, and we simulate 1200 epochs. The full loss goes down together to much lower values [Fig.~\ref{fig:FPE}(d)]. This translates into a high-quality PDF [Fig.~\ref{fig:FPE}(e)]. But most importantly, the presence of differential constraints provided high-quality derivatives plotted in Fig.~\ref{fig:FPE}(f). This paves the road to \emph{training models}, and not just learning from data, especially in cases where large datasets are not available or cannot be loaded efficiently.
We complete static FPE learning by sampling from optimal DQGM, based on the full loss. The originally-trained and extended 6- and 10-qubit sampling shown in  Fig.~\ref{fig:FPE}(g,h), showcases improvements offered by including the knowledge about the model and underlying SDE/PDE.
%%%
\begin{figure}
\begin{center}
\includegraphics[width=1.0\linewidth]{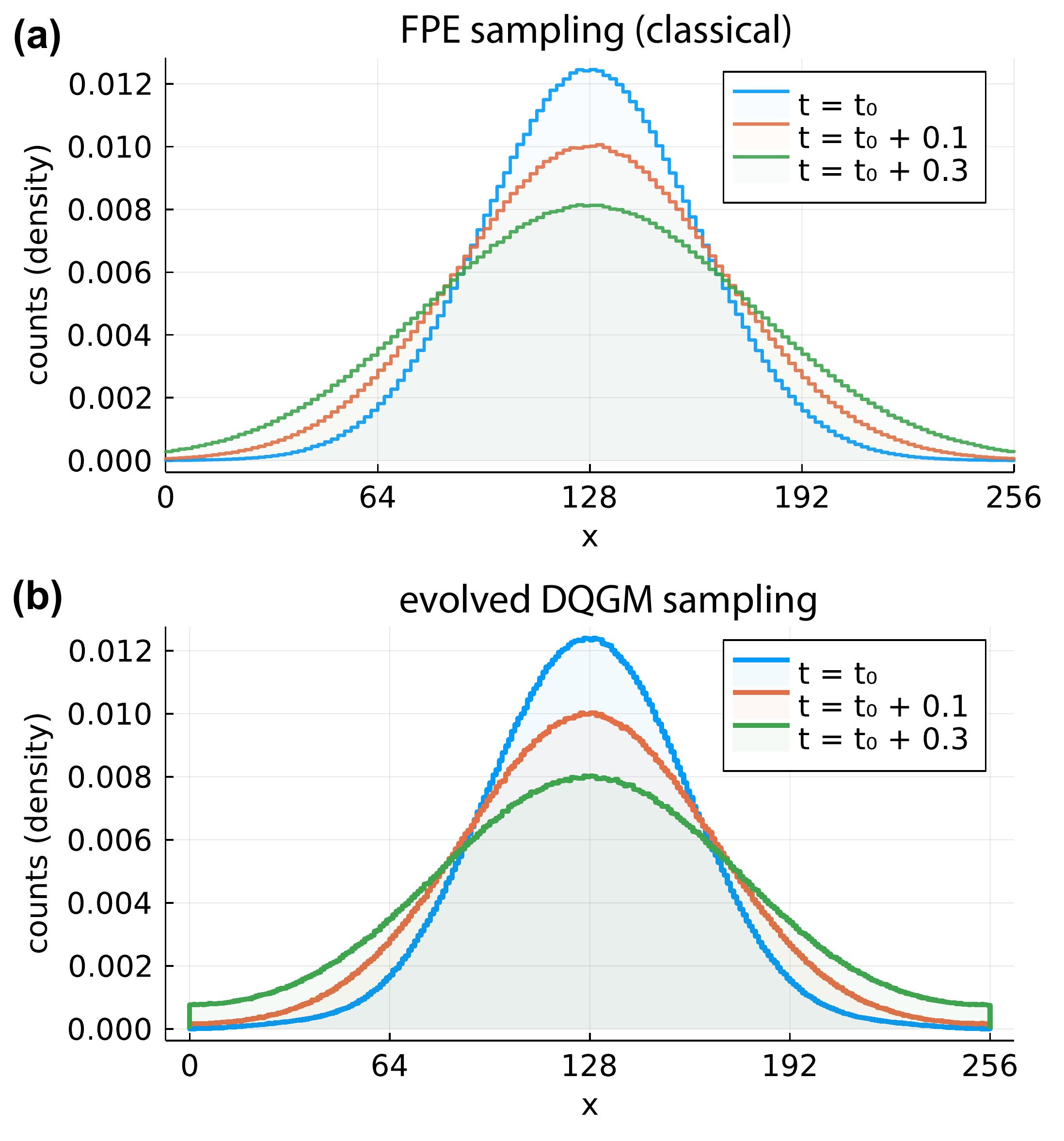}
\end{center}
\caption{\textbf{Time-dependent SDE sampling.} (a) Samples from classically evolved distribution at different time points ($t_0$, $t_0 + 0.1$, $t_0 + 0.3$). (b) Samples from time-evolved DQGM at the same three times obtained by evolving circuit parameters with the implicit time embedding.}
\label{fig:evolution}
\end{figure}
%%%

\subsection{Solving time-dependent Fokker-Planck equations}

Once the initial state is learnt and differential constraints are accounted for, we may ask an additional question: can we predict the trajectories of the stochastic process that lead to the steady state? To answer the question, let us first solve the problem using the conventional Euler-Maruyama technique \cite{Kloeden1992}. 

We set up an SDE solver for the OU process with increasing variance. For simplicity, we consider a process without mean reversion, setting $\mu = 32$, and a variance of $\sigma = 512$ as SDE/FPE parameters. We start from the delta function distribution at zero time, and learn the PDF at $t_0 = 0.144$ (in the units of inverse $\kappa$). At this point the distribution matches the variance of $64$, and continues to grow thereafter. The results from classical SDE sampling are shown in Fig.~\ref{fig:evolution}(a) for three different times being $t_0$, $t_0 + 0.1$, and $t_0 + 0.3$, chosen such that changes are significant. Next, we perform time-evolved simulation with the DQGM. We express the solution as DQGM at $t_0$ with a $w=2$ and $d=1$ circuit that performed well before, while choosing a variational circuit structure with real-amplitude states (layers of parametrized Y rotations and CZ gates). The training follows data-based loss and 500 epochs. Then, we assume the implicit time embedding, and update parameters of the model $\bm{\theta}_{\mathrm{opt},t_0}$ from the initial ones by evaluating the FPE operator and Jacobians. We use a simple Euler's scheme with $\Delta t = 0.001$ and three hundred steps. Note that this may lead to instability for longer propagation times, where Runge-Kutta and stencil-point methods are preferred. The histograms for time-evolved DQGM are shown in Fig.~\ref{fig:evolution}(b), where $10^7$ samples are used. We observe good agreement with classical sampling, and note that having a smooth model the sampling can further be extended to larger register sizes. We also note that explicit encoding may be beneficial for situations where we need to generalize in time. This will be a question for future research on the topic.
%%%
\begin{figure*}[t!]
\begin{center}
\includegraphics[width=1.0\linewidth]{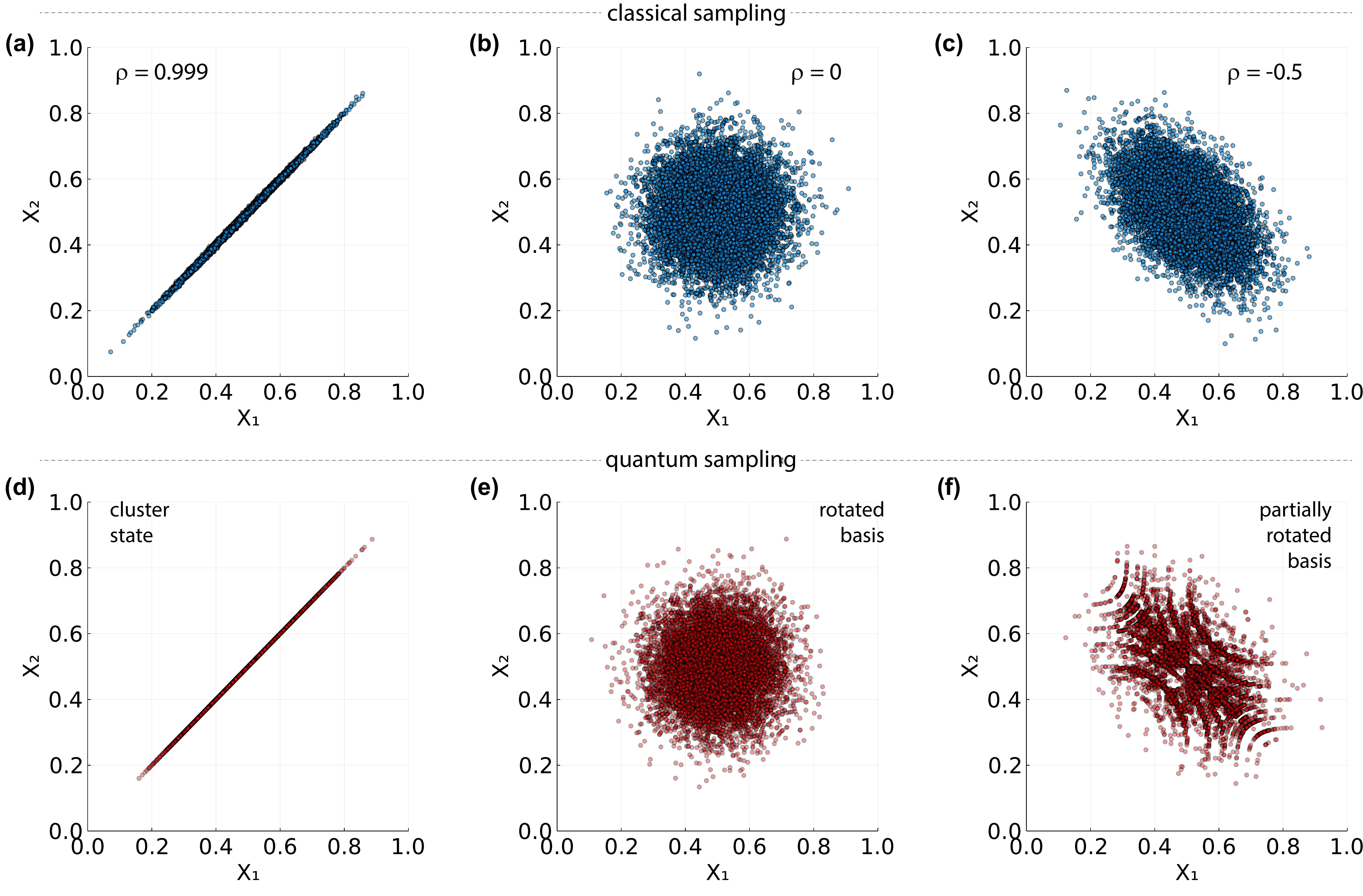}
\end{center}
\caption{\textbf{Classical and quantum multivariate sampling with normal copulas.} (a, b, c) Scatter plots for classical sampling of random variables $X_1$ and $X_2$ from the bivariate normal distribution. The probability density functions are centered at $0.5$, standard deviations are $0.1$, and correlation between variables is $\rho_{12} = \{ 0.999, 0.0, -0.5\}$ for (a), (b), and (c), respectively. $10^4$ samples are shown. (d) Scatter plot for quantum generative modelling from the maximally logically entangled copula transformed into normal samples and mimicking $\rho_{12} \rightarrow 1$ case. Here and below $N=12$ qubits are used for the full register, sample values are normalized to be in $[0,1]$ dividing by $2^{N/2}$, and $10^4$ are plotted. (e) Sampling from uncorrelated registers, where copula circuit has uncorrelated bases for the two registers. (f) Partially correlated copula transformed into bivariate samples that mimics negative $\rho_{12} = -0.5$ correlation.}
\label{fig:copula}
\end{figure*}
%%%

\subsection{Sampling from bivariate normal distributions}

Next, we study a pedagogical example of sampling from a multivariate distribution. We consider a bivariate normal distribution $p(x_1, x_2)$. This type of distribution can be fully characterized by its mean values for each stochastic variable $\mu_{1,2}$, their respective standard deviations $\sigma_{1,2}$, and importantly the correlation parameter $\rho_{12}$. Let us first analyse different examples using known classical procedures of inverse sampling which accounts for the covariance matrix. In Fig.~\ref{fig:copula}(a, b, c) we show three examples for classical bivariate sampling. The first example concerns highly-correlated samples $(X_1, X_2)$ with $\rho_{12} = 0.999$, each normally distributed with $\mu_{1,2} = 0.5$ and $\sigma_{1,2} = 0.1$ [Fig.~\ref{fig:copula}(a)]. One can think of financial processes with similar correlation at highly regulated markets, or for instance looking at EUR-DKK currency pair. Next, as a reference we show sampling from uncorrelated distribution with $\rho_{12} = 0$ [Fig.~\ref{fig:copula}(b)], which is equivalent to separate inverse sampling of $X_{1}$ and $X_2$, plotted together. The third example in Fig.~\ref{fig:copula}(c) concerns a negative correlation value of $\rho_{12} = -0.5$. This example is relevant in cases when significant but not absolute dependence of two processes is present.

We continue the analysis in the quantum domain using copula as a tool. First, we note that for multivariate normal processes the Gaussian copula PDF $c(\bm{z})$ can be expressed as
\begin{align}
\label{eq:copula}
c(\bm{z}) &= \frac{1}{\sqrt{1 - \rho_{12}^2}} \exp \bigg\{ \Big[2 \rho_{12} Q_1(z_1) Q_2(z_2) \\ \notag &- \rho_{12}^2 \big(Q_1^2(z_1) + Q_2^2(z_2) \big)\Big] / 2 (1 - \rho_{12}^2) \bigg\},
\end{align}
where $Q_j(z_j)$ are standard normal quantile functions for variables $j = 1,2$ expressed as (shifted) inverse error functions parametrized by $(\mu_j, \sigma_j)$ \cite{Shaw2008}. Now, let us look at the limiting cases. For $\rho_{12} = 0$ the copula PDF becomes the uniform distribution for both variables. In the limit of $\rho_{12} \rightarrow 1$ we get maximal correlations, such that it is given by the Dirac delta function, $c(z_1, z_2) \sim \delta(z_1 - z_2)$. For non-zero $\rho_{12}$ the structure is introduced, leading to preference of some samples over others. Using the described intuition from Gaussian copula, we note that perfect correlation of $\rho_{12} \rightarrow 1$ is readily modelled by a cluster state circuit with variational circuits being identities (here, it is easier to use the generalized QCBM picture for gaining the intuition). Once the copula circuit is set up, we perform mapping $Z_{1,2} \rightarrow X_{1,2}$ as described in Eq.~\eqref{eq:copula_sampling}, and present scatter plots for $(X_1, X_2)$.

The resulting samples in the multivariate data space are shown in Fig.~\ref{fig:copula}(d), resembling the highly correlated case discussed before. Next, the decorrelation circuit can be set up such that the measurement for $Z_1$ and $Z_2$ are performed in different bases, for instance acting with Hadamards on the first register. The corresponding sampling is shown in Fig.~\ref{fig:copula}(e), mimicking $\rho_{12} = 0$ case. Finally, by employing single qubit rotations on the first register in X and Y basis the partial correlation can be reproduced [see Fig.~\ref{fig:copula}(f)]. 

We note that the present study shows only the first steps in understanding multidimensional correlated sampling from quantum circuits. However, using the developed tools and combining with knowledge of stochastic processes may improve this understanding ever further.

\section{Conclusion}
We developed protocols for efficiently training differentiable quantum generative models, which we refer to as DQGM. Separating training and sampling stages, we train circuits in the latent space as a feature map encoded differentiable circuit, and sample the optimized circuit with additional (fixed) basis transformation. On a technical side, we introduced the phase feature map, analyzed its properties, and developed frequency-taming techniques that include qubit-wise training and feature map sparsification. For numerical simulations, we benchmark the approach against QCBM and show how samples from propagated stochastic differential equations can be accessed by solving a Fokker-Planck equation on a quantum computer. Our approach also sheds light on a path to multidimensional generative modelling based on copulas, where qubit registers are explicitly correlated via a (fixed) entangling layer. In this case quantum computers can offer advantage as efficient samplers, which perform complex inverse transform sampling enabled by fundamental laws of quantum mechanics.

\textit{Ethics declaration.} A patent application for the method described in this manuscript has been submitted by Pasqal.

%---------------------------------
% The bibliography
%---------------------------------

%\begin{flushleft}\textbf{AUTHOR CONTRIBUTIONS}\end{flushleft}\vspace{-2mm}
%
%\noindent \small{A.\,E.\,P. performed the calculations, described the results, and analyzed the qGAN performance. O.\,K. proposed the original idea and overseen the project together with V.\,E.\,E. All authors contributed to writing the manuscript and analyzing the results.}\vspace{1mm}

%\begin{flushleft}\textbf{COMPETING INTERESTS}\end{flushleft}\vspace{-2mm}
%
%\noindent \small{The authors declare that there are no competing interests.}\vspace{1mm}

%-------------------------------
%\clearpage

%\newpage

%\appendix

%\section{Methods}

%In the Appendices we provide the background search, describe the procedure for time-dependent PDE solving, and give training details.\vspace{2mm}

%\subsection{Background search}

%\noindent For the background search, mostly related to the patent applications, we can highlight the following studies. First, in [..] the authors considered...

%\subsection{Quantum Fourier transform}

%\noindent As a simplest transformation circuit we employ an inverse quantum Fourier transform. For the numerical simulations we use the following circuit...

\end{document}